\documentclass[pdflatex,sn-mathphys-num]{sn-jnl}
\usepackage{graphicx}
\usepackage{multirow}
\usepackage{amsmath,amssymb,amsfonts}
\usepackage{array}
\usepackage{booktabs}
\usepackage{algorithm}
\usepackage{algpseudocode}
\usepackage{siunitx}
\usepackage[table]{xcolor}
\usepackage{colortbl}
\usepackage{diagbox}
\usepackage{booktabs}
\usepackage{makecell}
\usepackage{float}
\usepackage{placeins}
\usepackage{url}
\usepackage{verbatim}
\usepackage{subfig}
\newcolumntype{C}[1]{>{\centering\arraybackslash}p{#1}}
\usepackage{textcomp}
\usepackage{hhline}
\usepackage{lineno}

\begin{document}

\title[Short title]{Design and Performance Evaluation of a Quad-panel Dual-isotope PET Imaging System}

\articletype{Original Research Article}

\author[1,2]{\fnm{Yu} \sur{Shi}}
\equalcont{These authors contributed equally to this work.}
\author[2]{\fnm{Tengjiao} \sur{Wang}}
\equalcont{These authors contributed equally to this work.}
\author[2]{\fnm{Sen} \sur{Yang}}
\author[2]{\fnm{Bo} \sur{Wen}}
\author[3]{\fnm{Meng} \sur{Niu}}
\author[3]{\fnm{Yirong} \sur{Wang}}
\author[3]{\fnm{Risheng} \sur{Yang}}
\author[2]{\fnm{Yanyun} \sur{Liu}}
\author[1]{\fnm{Wei} \sur{Ding}}
\author[2]{\fnm{Xiangyu} \sur{Liu}}
\author*[3]{\fnm{Fei} \sur{Kang}}\email{fmmukangf@fmmu.edu.cn}
\author*[2]{\fnm{Shouping} \sur{Zhu}}\email{spzhu@xidian.edu.cn}

\affil[1]{\orgdiv{School of Electrical Engineering and Automation}, \orgname{Suzhou University of Technology}, \orgaddress{\city{Suzhou}, \postcode{215500}, \state{Jiangsu}, \country{China}}}
\affil[2]{\orgdiv{School of Life Science and Technology}, \orgname{Xidian University \& Engineering Research Center of Molecular and Neuro Imaging, Ministry of Education}, \orgaddress{\city{Xi'an}, \postcode{710126}, \state{Shaanxi}, \country{China}}}
\affil[3]{\orgdiv{Department of Nuclear Medicine}, \orgname{Xijing Hospital, Fourth Military Medical University}, \orgaddress{\city{Xi'an}, \postcode{710032}, \state{Shaanxi}, \country{China}}}

\abstract{
\textbf{Background:}
Simultaneous dual-isotope PET can provide complementary molecular information from two radiotracers within a single acquisition.
However, prompt-gamma-based separation is challenged by the low statistics of triple coincidences and by the limited flexibility of conventional coincidence logic for multi-energy event extraction. We designed and evaluated a compact quad-panel dual-isotope PET system without dedicated prompt gamma detectors.

\textbf{Methods:}
The system comprised four orthogonally arranged detector panels based on LYSO crystal/silicon photomultiplier modules, each providing an effective area of 107.0 mm $\times$ 145.5 mm. GPU-accelerated software coincidence processing extracted double- and prompt-gamma-tagged triple-coincidence events from raw single-photon data. Triple-coincidence images were reconstructed with correlated-random-triple correction using the consistent delayed expectation-maximization method, followed by image-domain isotope separation using a spatially varying scaling factor. System performance was evaluated using phantom and \textit{in vivo} studies, with emphasis on $^{68}$Ga/$^{18}$F imaging because of the low prompt gamma yield of $^{68}$Ga (1.2\%).

\textbf{Results:}
The system achieved energy resolutions of 15.7\%, 12.9\% and 10.2\% at 511, 1077, and 1275 keV, respectively, and a coincidence timing resolution of 1.35 ns.
In the $^{68}$Ga Derenzo phantom, 1.6-mm rods were resolved in double-coincidence images, with 1.2-mm rods partially distinguishable, while 1.6-mm rods remained discernible in triple-coincidence images.
The mean prompt gamma detection efficiencies along the three axes ranged from 8.19\% to 9.31\% @ 1275 keV. The system sensitivity was 1.12\% for double coincidences with an energy window of 350--650 keV, and 0.068\% for triple coincidences after correlated-random-triple correction with a 350--650 \& 900--1400 keV dual-energy window. The scatter fraction was 27.81\%, and the peak noise-equivalent count rates reached 83 kcps at 9.38 MBq for doubles and 106 cps at 21.42 MBq for triples. Phantom and \textit{in vivo} studies demonstrated successful separation of $^{18}$F/$^{68}$Ga and $^{18}$F/$^{124}$I tracer distributions.

\textbf{Conclusions:}
The proposed quad-panel PET system enables simultaneous dual-isotope imaging using standard PET detector modules and software-based coincidence processing. By avoiding additional prompt gamma detectors while maintaining high-resolution and quantitative imaging performance, the system provides a practical preclinical platform for multi-tracer PET studies involving non-pure positron emitters with low prompt gamma yields.

\textbf{Clinical trial number:}
Not applicable.
}


\maketitle

\section{Background}
Positron Emission Tomography (PET) is a powerful molecular imaging modality in both clinical diagnostics and biomedical research, enabling non-invasive visualization and quantification of physiological and biochemical processes \textit{in vivo}. Conventional PET imaging is typically performed with a single radiotracer, which provides information on a specific molecular target or metabolic pathway. For instance, $^{18}$F-fluorodeoxyglucose ($^{18}$F-FDG) is widely applied to assess glucose metabolism\cite{1}, whereas $^{68}$Ga-labeled tracers are used to characterize receptor-specific tumor biology, such as $^{68}$Ga-DOTA-conjugated somatostatin receptor analogues for neuroendocrine tumors (NETs)\cite{3} and $^{68}$Ga-PSMA for prostate cancer\cite{2}. However, single-tracer PET may be insufficient to characterize complex disease biology, where metabolic activity, receptor expression, perfusion, hypoxia, or stromal remodeling may provide complementary information\cite{4}. Dual-tracer or dual-isotope PET has therefore attracted increasing interest because it enables complementary characterization of distinct molecular pathways, thereby potentially enhancing diagnostic sensitivity and specificity. For example, combined assessment with $^{18}$F-FDG and $^{68}$Ga-FAPI has been reported to improve multiple myeloma evaluation by revealing both metabolic activity and stromal alterations\cite{5}.

In clinical practice, dual-isotope PET is commonly implemented via sequential scanning, in which two radiotracers are administered on separate days and followed by multiple PET/CT acquisitions\cite{6}. Although straightforward, this protocol increases examination time and CT-related radiation exposure, and may introduce spatial misregistration and physiological variability between acquisitions, thereby complicating image interpretation and quantitative comparison\cite{7}. To overcome these limitations, several approaches have been explored for single-session dual-tracer PET. One early approach exploits the spatial separation of tracer biodistributions, such as the combined use of $^{18}$F-FDG and $^{18}$F-NaF in organs with non-overlapping uptake patterns\cite{8}. While operationally simple, this method is limited to radiotracer pairs with sufficiently distinct biodistributions. Alternatively, dual-tracer separation based on metabolic kinetic modeling has been proposed using dynamic imaging protocols\cite{9_,10_}. However, these techniques require prolonged acquisitions ($>$45 min) and are sensitive to model assumptions, limiting their clinical feasibility. More recently, deep-learning-based methods have been developed to separate dual-tracer PET signals by learning tracer-specific features directly from data~\cite{9,10,11}. Although these methods can reduce the need for time-separated injections and explicit kinetic modeling, they generally still rely on multi-frame dynamic acquisitions, representative training data and extended scan durations.

A fundamentally different strategy is prompt-gamma-based simultaneous dual-isotope imaging, which exploits prompt gamma photons emitted in temporal coincidence with positron decay in non-pure positron emitters. By identifying triple-coincidence events consisting of two 511-keV annihilation photons and one additional prompt gamma photon, isotope-specific information can be retrieved from a static acquisition without requiring dynamic scanning. In 2011, Andreyev and Celler first demonstrated the feasibility of this concept through Monte Carlo simulations, showing that dual-isotope combinations such as $^{18}$F/$^{22}$Na and $^{18}$F/$^{60}$Cu could be separated with an error below 5\% using triple coincidence identification\cite{12}. However, most commercial PET systems rely on hardware-based coincidence logic with predefined energy windows, which is not designed for flexible multi-energy triple-coincidence processing. To address this limitation, Gonzalez et al.\cite{13} and Miyaoka et al.\cite{14} incorporated dedicated prompt gamma detectors into PET systems and demonstrated the separation of $^{68}$Ge/$^{22}$Na in phantom studies. In 2017, Fukuchi et al. further improved system flexibility and detection efficiency by developing a multi-isotope PET (MI-PET) system, in which two BGO prompt gamma detector rings were placed around a central PET scanner\cite{15}. This system enabled $^{18}$F/$^{22}$Na separation in small animals. More recently, Pratt et al. demonstrated a post-processing approach for triple-coincidence extraction using commercial PET hardware, thereby avoiding hardware modifications. Nevertheless, this approach remains constrained by the available hardware coincidence logic and is mainly applicable to non-pure positron emitters with prompt gamma emissions close to the 511-keV annihilation window, such as $^{124}$I\cite{16}.

Prompt gamma-based imaging of multiple positron emitters has been recognized as a promising direction for next-generation diagnostic PET technologies\cite{2025TRPMS}. However, practical implementation remains challenging because triple-coincidence events are intrinsically scarce and strongly affected by isotope-dependent physical factors, including prompt gamma yield, prompt gamma energy, positron range, random coincidences and dead time~\cite{27}. Therefore, flexible coincidence processing, robust random-triple correction and comprehensive system characterization are essential for practical prompt-gamma-enabled dual-isotope PET. 
Our group has previously investigated algorithms for prompt-gamma-based dual-isotope PET imaging~\cite{Shi2022PromptSpectrum, 22}.
The present study focuses on the overall design and systematic performance evaluation of a modular quad-panel dual-isotope PET system that enables prompt gamma detection without dedicated prompt gamma detectors. The proposed system acquires raw single-photon data and employs GPU-accelerated software-based coincidence processing to flexibly extract both double- and triple-coincidence events. Subsequently, double- and triple-coincidence images are reconstructed using expectation-maximization (EM) based algorithms, followed by image-domain isotope separation. Considering the extremely low prompt gamma emission probability of $^{68}$Ga (1.2\%), this study focuses on $^{68}$Ga/$^{18}$F dual-isotope imaging to evaluate the system’s robustness and generalizability. A series of phantom and small-animal experiments were conducted to characterize system performance and validate dual-isotope separation. This study demonstrates the feasibility of prompt-gamma-enabled dual-isotope PET using standard detector modules alone, while maintaining high-resolution and quantitative imaging performance.

The remainder of this paper is organized as follows. Section 2 describes the system architecture, coincidence processing, correction and reconstruction methods, isotope separation strategy, and experimental protocols. Section 3 presents the system performance and dual-isotope imaging results. Sections 4 and 5 provide the discussion and conclusions, respectively.

\begin{figure}[!bp]
\centering
\subfloat[]{\includegraphics[height=1.4in]{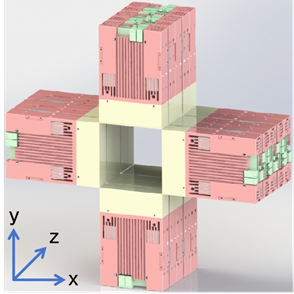}%
\label{fig_1_a}}
\hfil
\subfloat[]{\includegraphics[height=1.4in]{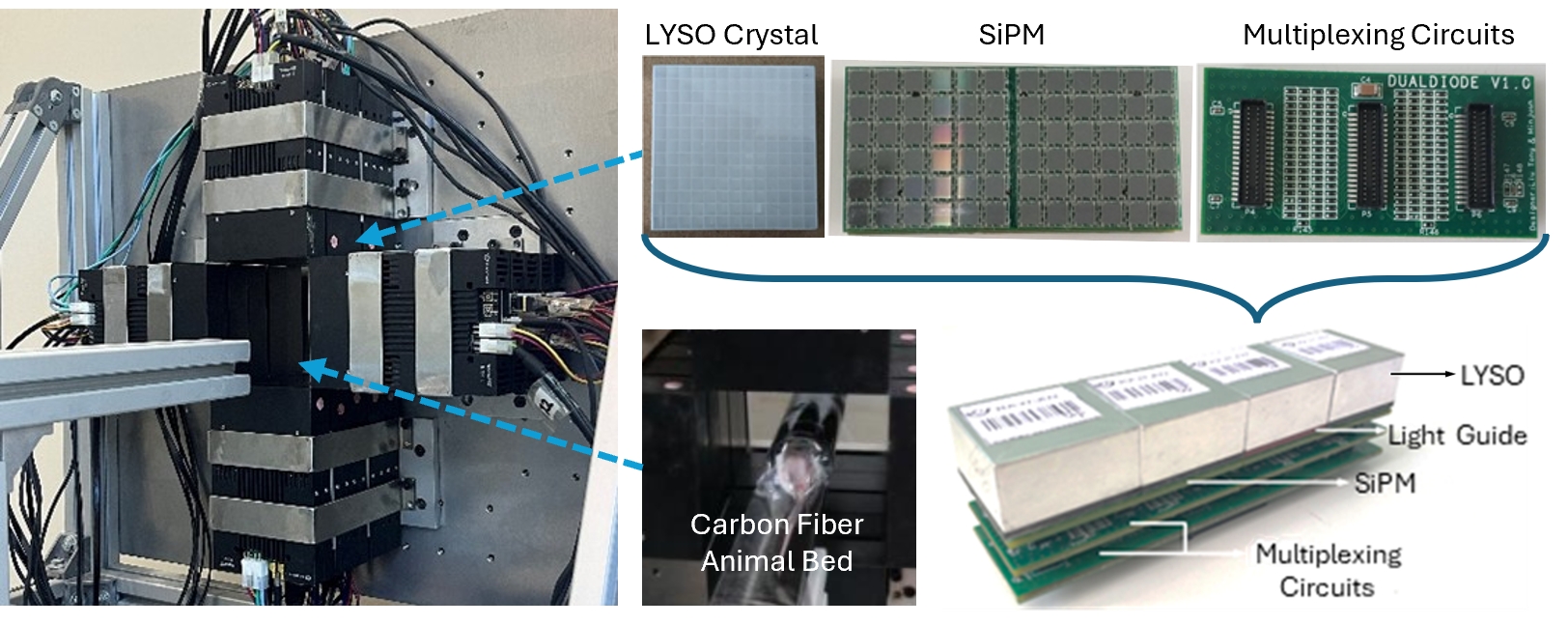}%
\label{fig_1_b}}
\caption{Structure of the quad-panel dual-isotope PET system. (a) Simulated configuration. (b) Physical system.}
\label{fig_1}
\end{figure}

\section{Materials and methods}
\subsection{System Architecture}
The proposed dual-isotope PET system is constructed upon modular digital detector units, namely BDM II modules\cite{18,19}, developed by Raycan Technology Co., Ltd. (Suzhou, China). Each module employs a fully digital multi-voltage threshold (MVT) architecture and integrates a 13$\times$52 array of LYSO crystals, each measuring $1.89 \times 1.89 \times 13.00~\mathrm{mm}^3$, coupled to SiPM arrays through acrylic light guides. The BDM modules provide an energy resolution better than 15\% at 511 keV, and a maximum counting rate of approximately 2.0 Mcps. Five BDM modules are stacked in parallel to form one detector panel, and four such panels are arranged orthogonally to form a box-shaped configuration, defining the system's imaging field of view (FOV). Each panel has an effective detection area of $107.0~\mathrm{mm} \times 145.5~\mathrm{mm}$, and the distance between opposing detector panels is $109~\mathrm{mm}$. Two 2-mm inter-module gaps are present in each panel owing to the space required for the Ethernet interfaces. The simulated configuration and physical prototype of the system are shown in Fig. \ref{fig_1_a} and Fig. \ref{fig_1_b}, respectively. The $Z$ axis is defined along the axial direction, while the $X$ and $Y$ axes are aligned with the transverse plane of the FOV. All 20 BDMs are daisy-chained for power distribution and synchronized through a central clock board, ensuring simultaneous global resets and consistent timestamp alignment across modules. The acquired single-photon event data are packetized and transmitted to a host workstation via 10-gigabit optical fibers for real-time or offline processing, including software-based coincidence detection and image reconstruction.

\subsection{Coincidence Detection}
Following the acquisition of single-photon event data, coincidence processing is performed to identify both double- and triple- coincidences based on the distinct physical properties of pure and non-pure emitters. A double coincidence refers to the detection of a pair of 511 keV annihilation photons originating from a positron-emitting decay, and is applicable to all positron emitters. In contrast, a triple coincidence consists of two annihilation photons and an additional prompt gamma photon, and is characteristic of non-pure positron emitters that emit prompt gamma photons during nuclear de-excitation.

\begin{figure}[htbp]
\centering
\includegraphics[width=0.9\textwidth]{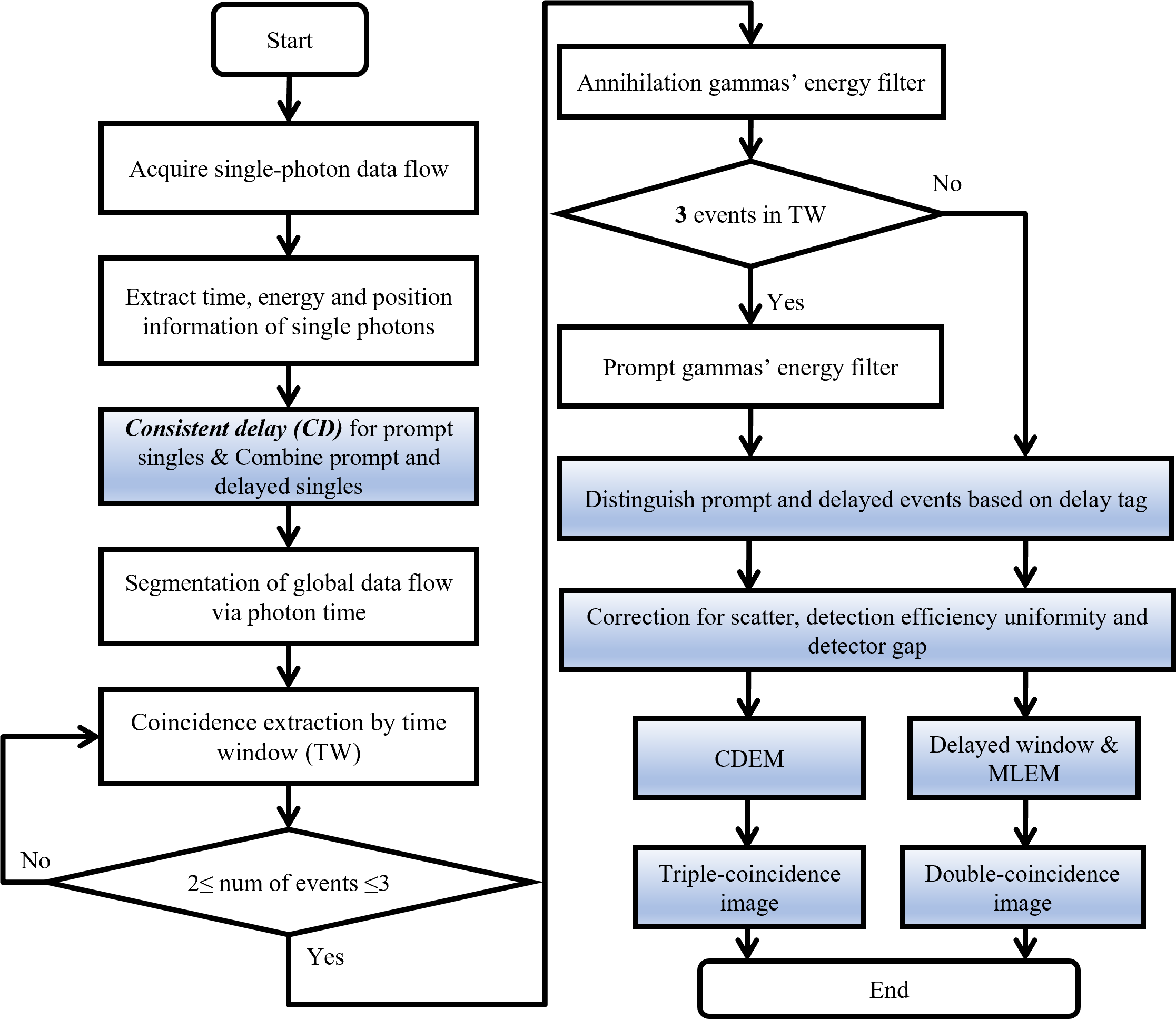}
\caption{Coincidence processing, correction and reconstruction workflow for double- and triple-coincidence events. The blue-shaded blocks indicate correction and reconstruction steps.}
\label{fig_2}
\end{figure}

The overall processing workflow is illustrated in Fig. \ref{fig_2}. Considering the approximately 15\% energy resolution at 511 keV achieved by the BDM modules, an energy window of 350-650 keV is applied to extract annihilation photons. A coincidence time window of 6 ns is employed to identify temporally correlated photon pairs. Events satisfying both the energy and timing criteria are registered as valid double coincidences and used to construct lines of response (LORs). For triple-coincidence processing, a third photon corresponding to a prompt gamma is required in coincidence with the annihilation photon pair. In this study, $^{68}$Ga serves as the non-pure positron emitter, which emits a 1077-keV prompt gamma photon with a probability of 1.2\%. To extract such triple events, an additional energy window of 900--1400 keV is applied for prompt gamma detection, while the same 6 ns time window is maintained. The annihilation photon pairs within these triple-coincidence events are then used to construct triple-coincidence LORs. All coincidence processing was implemented using a GPU-accelerated software-based coincidence processing framework (GPU-SCP), developed in our previous work\cite{20}. This framework enables flexible multi-energy coincidence filtering and high-throughput data processing, making it suitable for dual-isotope PET applications.

\subsection{Correction and Reconstruction}
Corrections are applied to both double and triple coincidences, including random correction, scatter correction, normalization, and compensation for inter-module gaps. Except for random correction, the same correction procedures are applied to both coincidence modes. 
Owing to the inherently low count statistics of triples, conventional random pre-correction methods, such as the delayed window methods, suffer from significant noise problems and are ineffective in addressing negative values in projections\cite{21}. In contrast, iterative estimation based on a joint Poisson model produces more stable and variance-reduced correction. Therefore, the previously proposed Consistent Delayed Expectation-Maximization (CDEM) algorithm\cite{22} is employed to correct for random triples with robust estimation.

Among various types of random triples, correlated random triples (CRTs) are predominant\cite{23}. A CRT consists of a true annihilation-photon pair originating from the same positron annihilation event and an unrelated gamma photon from a different decay. Because the annihilation photon pair in a CRT remains temporally correlated, conventional delayed-window techniques may lead to biased estimates when independent delays are applied to different channels, as this operation disrupts the intrinsic correlation of the true double-coincidence component.

To realize the effective estimation of random triples, the CDEM framework treats each decay event as the fundamental unit rather than treating photons independently. Specifically, photons are delayed by a consistent interval across all channels, thereby preserving the internal temporal correlation within each decay while exploiting the statistical independence between different decays to correct random triples. After applying the coincidence time window to the merged prompt and delayed singles, random triples $n_d$ are extracted via delay tag.

Both true and random triple coincidences are modeled as Poisson processes. The observed prompt triple counts are treated as the superposition of true and random events, and are incorporated into a joint Poisson likelihood function. The mathematical expression of the above description is as
\begin{equation}
\label{eqn_1}
\begin{aligned}
\left\{ \begin{array}{l}
{n_p}(i) \sim \mathrm{Poisson}\left(\sum\limits_j a_{ij}\left( {f_t(j) + f_r(j)} \right)  \right),\\
n_d(i) \sim \mathrm{Poisson}\left( \sum\limits_j {a_{ij}f_r(j)}  \right),
\end{array} \right.
\end{aligned}
\end{equation}
where $n_p(i)$ represents the number of prompt triples detected by the $i$th LOR, $n_d(i)$ denotes the number of random triples extracted by the consistent delayed window on the $i$th LOR, and $a_{ij}$ is the system response matrix. The variables $f_t(j)$ and $f_r(j)$ denote the spatial distributions of true triples from the non-pure emitter and random triples, respectively.

The joint log-likelihood of the prompt and delayed triple data is formulated as follows. Furthermore, total variation (TV) is also included in the objective function to suppress noise.
\begin{equation}
\label{eqn_2}
\begin{aligned}
L(f_t(j),  f_r(j)) = 
&\sum_{i = 1}^I \Bigg(
    - \sum_j a_{ij}(f_t(j) + f_r(j)) + n_p(i) \ln \bigg( \sum_j a_{ij}(f_t(j) + f_r(j))\bigg)
\Bigg)\\
& + \sum_{i = 1}^I \Bigg(
    - \sum_j a_{ij} f_r(j)
    + n_d(i) \ln \bigg( \sum_j a_{ij} f_r(j) \bigg)
\Bigg)\\
&-\beta \operatorname{TV}(f_t) - \beta \operatorname{TV}(f_r).
\end{aligned}
\end{equation}
The objective function in Eq.~\ref{eqn_2} is optimized using an EM-based iterative algorithm. The update equations are given by
\begin{equation}
\label{eqn_3}
\left\{
\begin{aligned}
f_t^{k + 1}(j) = & \frac{f_t^k(j)}{A_j + \beta U_t^k(j)}\sum_i \frac{a_{ij}n_p(i)}{\sum_{j^{'}} a_{ij^{'}}(f_t^k(j^{'})+f_r^k(j^{'}))}\\
f_r^{k + 1}(j) = &\frac{f_r^{k}(j)}{2{A_j} + \beta U_r^k(j)}\sum_i \bigg(\frac{{a_{ij}n_p}(i)}{\sum_{j^{'}} a_{ij'}(f_t^k(j^{'})+f_r^k(j^{'}))} \\
& +\frac{a_{ij}n_d(i)}{\sum_{j^{'}}a_{ij^{'}}f_r^k(j^{'})}\bigg)\\
\end{aligned}
\right.
\end{equation}
where $U_t^k(j)$ and $U_r^k(j)$ denote the partial derivatives
of $\operatorname{TV}(f_t)$ and $\operatorname{TV}(f_r)$ with
respect to $f_t(j)$ and $f_r(j)$, respectively, evaluated at the
current iteration. The parameter $\beta$ is the shared regularization weight, and $A_j$ is $\sum_ia_{ij}$.

Other correction schemes are performed in the projection domain before EM-based reconstruction. Specifically, an energy-spectrum-based method is used for scatter correction, and inverse-distance-weighted interpolation is used to compensate for inter-module gaps. The detailed implementation of these correction procedures has been described in our previously developed breast PET system\cite{24}.

\subsection{Separation}
Following reconstruction of the double- and triple-coincidence images using MLEM and CDEM, respectively, isotope separation is performed in the image domain. The purpose of this step is to decouple the distributions of the pure and non-pure positron emitters from the superposed double-coincidence image. Specifically, a spatially varying scaling factor $\alpha(j)$\cite{12}, determined by the system geometry and prompt gamma emission probability, is used to map the triple-coincidence image to the double-coincidence image domain. Then the mapped non-pure-emitter component is subtracted from the double-coincidence image to estimate the pure-emitter distribution:
\begin{equation}
\label{eqn_4}
\left\{
\begin{aligned}
f_{\mathrm{npure}}(j) &= \alpha(j)\, f_t(j),\\
f_{\mathrm{pure}}(j) &= f_{\mathrm{double}}(j)-f_{\mathrm{npure}}(j),
\end{aligned}
\right.
\qquad
\text{where}\quad
\alpha(j)=\frac{\theta \lambda(E)\, n_{\mathrm{dbl}}(j)}{n_{\mathrm{tri}}(j)}.
\end{equation}
Here, $f_{\mathrm{double}}(j)$ denotes the reconstructed double-coincidence image, $f_{\mathrm{npure}}(j)$ denotes the estimated non-pure-emitter contribution in the double-coincidence image domain, and $f_{\mathrm{pure}}(j)$ denotes the estimated pure-emitter distribution. The terms $n_{\mathrm{dbl}}(j)$ and $n_{\mathrm{tri}}(j)$ represent the numbers of double and triple coincidences extracted from a $^{22}$Na calibration scan at position $j$, respectively. 

A $^{22}$Na point source serves as a suitable calibration source for determining the scaling factor because of its nearly unity prompt gamma emission probability and negligible random-triple rate\cite{23}. The factors $\theta$ and $\lambda(E)$ are dimensionless correction ratios relative to the $^{22}$Na calibration. Specifically, $\theta$ is the ratio of the positron-associated prompt gamma yield of $^{22}$Na to that of the target radionuclide, whereas $\lambda(E)$ is the ratio of their effective prompt gamma detection efficiencies under the respective energy-window and event-selection conditions. These factors adapt the $^{22}$Na-derived scaling factor to the target non-pure positron emitter.
Therefore, a $^{22}$Na point source was scanned at different positions across the FOV to estimate the spatially varying scaling factor $\alpha(j)$.

\subsection{Experimental Setup}
\subsubsection{Energy and Timing Resolution}
Energy resolution was measured to characterize the ability of the system to discriminate incident gamma photons with different energies. In this study, point sources of $^{68}$Ga and $^{22}$Na were used to evaluate system energy performance at the 511-keV annihilation photopeak and at the prompt gamma photopeaks of 1077 keV and 1275 keV, respectively. For each detector crystal, the energy spectrum was extracted from the raw single-photon data, and the energy resolution was calculated as the full width at half maximum (FWHM) of the corresponding photopeak divided by the peak energy. Data acquisition was performed over 10 minutes for each source, and the average energy resolution was calculated across all detector elements.

Coincidence timing resolution was evaluated by calculating the distribution of time differences between coincidence photon pairs. A Gaussian function was fitted to the timing-difference spectrum, and the FWHM of the fitted curve was reported as the system’s coincidence timing resolution. The same acquisition setup as that used for the energy-resolution measurement was adopted to ensure consistency.

\subsubsection{Spatial Resolution}
Spatial resolution reflects the system’s ability to distinguish closely spaced structures within the imaging field. In the context of dual-isotope PET, spatial resolution was evaluated separately for double- and triple-coincidence imaging.

\begin{figure}[htbp]
\centering
\subfloat[]{\includegraphics[width=0.4\textwidth]{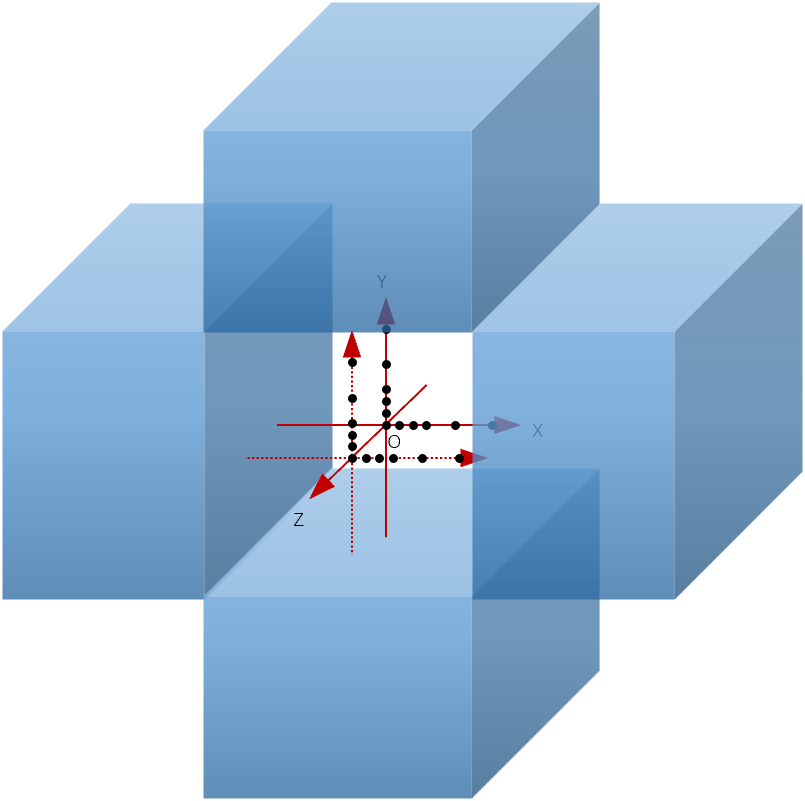}%
\label{fig3a}}
\hfil
\subfloat[]{\includegraphics[width=0.4\textwidth]{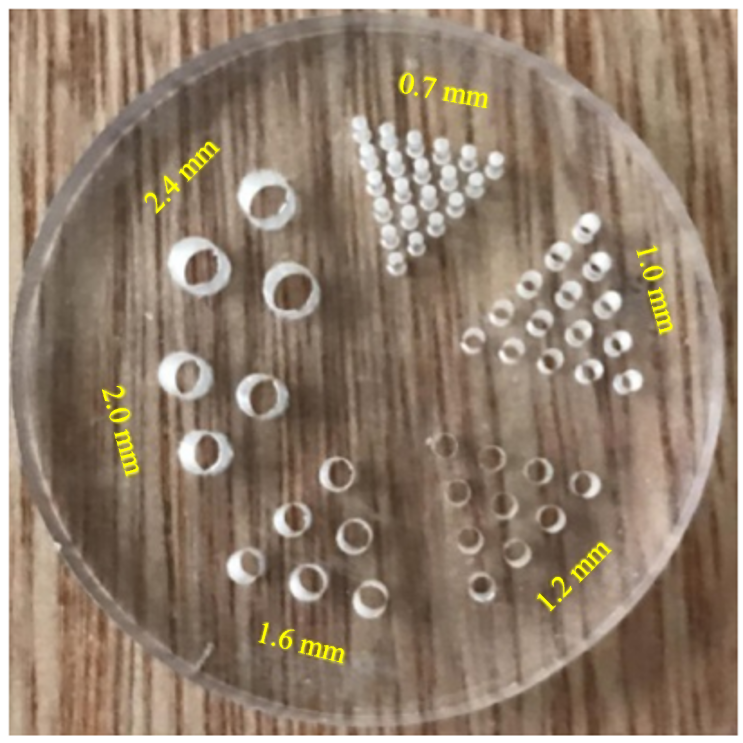}%
\label{fig3b}}
\caption{Phantoms used to measure the spatial resolution of the system. (a) $^{22}$Na point source distribution. (b) Small Derenzo phantom.}
\label{fig3}
\end{figure}

For double coincidences, spatial resolution was measured using conventional PET reconstruction based on detected annihilation photon pairs. Following the NEMA NU 4-2008 protocol, a $^{22}$Na point source was placed at multiple offset positions within the FOV, as shown in Fig.\ref{fig3a}. At each position, a 3D image was reconstructed from the double coincidence data. Line profiles were drawn through the peak of the reconstructed point spread function (PSF), and both the FWHM and full width at tenth maximum (FWTM) were measured along the $X$, $Y$, and $Z$ directions. The same procedure was applied to triple coincidences acquired at the same source positions, enabling a direct comparison of spatial resolution between the two modalities.

To further validate spatial resolution, a small Derenzo phantom was filled with 98 $\mu$Ci of $^{68}$Ga solution and scanned for 15 minutes. The phantom contained cylindrical rods with diameters ranging from 0.7 mm to 2.4 mm, as shown in Fig.\ref{fig3b}. Images were reconstructed using 100 iterations of MLEM (double) and CDEM (triple), allowing visual assessment of the minimum resolvable structures.

\subsubsection{Prompt Gamma Detection Efficiency}
Prompt gamma detection efficiency reflects the system’s ability to detect high-energy prompt gamma photons. According to the definition proposed by Fukuchi et al.\cite{15}, the prompt gamma detection efficiency $P_g$ was calculated as
\begin{equation}
    \label{eqn_5}
    P_g = \frac{P_t}{P_d},
\end{equation}
where $P_d$ is the double coincidence detection efficiency for annihilation photons, and $P_t$ denotes the triple-coincidence detection efficiency.
Since PET system sensitivity is highly dependent on the source position within the FOV, $P_g$ was measured at multiple spatial locations using a 12-$\mu$Ci $^{22}$Na point source. The $^{22}$Na source was selected because of its nearly unity prompt gamma emission probability of 99.9\%, allowing the prompt gamma detection efficiency to be estimated without additional correction for the emission branching ratio. For the $^{22}$Na point-source measurements, double-coincidence events were extracted using a 350--650-keV energy window for annihilation photons, whereas triple-coincidence events were extracted by requiring two annihilation photons within 350--650 keV and an additional prompt gamma photon within 1000--1500 keV. The same 6-ns coincidence time window was used for both double- and triple-coincidence extraction.
The point source was placed along the $X$, $Y$, and $Z$ axes at offsets of 0, 10, 20, 30, and 40 mm from the center of the FOV. At each position, data were acquired for 10 minutes. The numbers of detected double and triple coincidences were recorded to calculate $P_g$ according to Eq.~\ref{eqn_5}.

\subsubsection{System Sensitivity}
System sensitivity was measured in accordance with Chinese Industry Standard YY/T 1835-2022. $^{68}$Ga was used to evaluate system sensitivity because it is the primary non-pure positron emitter investigated in this work, with a prompt gamma emission probability of 1.2\%. A line-source phantom filled with $^{68}$Ga solution
, with an initial activity of 176 $\mu$Ci (6.5 MBq), was positioned at the axial center of the FOV and scanned for 15 minutes. A second acquisition was performed using a non-radioactive phantom for background subtraction.

For the proposed dual-isotope PET system, sensitivity was evaluated separately for double- and triple-coincidence events. Double-coincidence sensitivity was calculated using single-slice rebinning (SSRB). The sensitivity of the $i$th axial slice and the total system sensitivity were given by:
\begin{equation}
    \label{eqn_6}
    \left\{
    \begin{aligned}
        S_i = \frac{C_i-B_i}{A_0b_{\beta^+}}\\
        S_{total} = \sum_i{S_i}
    \end{aligned}
    \right.
\end{equation}
where $C_i$ is the detected double-coincidence count rate of the $i$th slice, $B_i$ is the corresponding background count rate, $A_0$ is the total source activity, and $b_{\beta^+}$ is the positron emission branching ratio. The total system sensitivity was obtained by summing $S_i$ across all axial slices.

For triple coincidences, the joint branching ratio for positron emission accompanied by a prompt gamma photon was incorporated into the sensitivity calculation:
\begin{equation}
    \label{eqn_7}
    S_i^{(T)} = \frac{C_i^{(T)}-B_i^{(T)}}{A_0f}
\end{equation}
where $C_i^{(T)}$ and $B_i^{(T)}$ represent the triple-coincidence count rate and background count rate of the $i$th slice, respectively. $f$ is the joint branching ratio for positron emission accompanied by the 1077-keV prompt gamma photon. For $^{68}$Ga, $b_{\beta^+}$ and $f$ were taken as 88.9\% and 1.2\%, respectively.
The total triple-coincidence sensitivity was calculated by summing $S_i^{(T)}$ across axial slices.

\subsubsection{Scatter Fraction and Count Rate Performance}
\begin{figure}[htbp]
\centering
\subfloat[]{\includegraphics[width=0.5\textwidth]{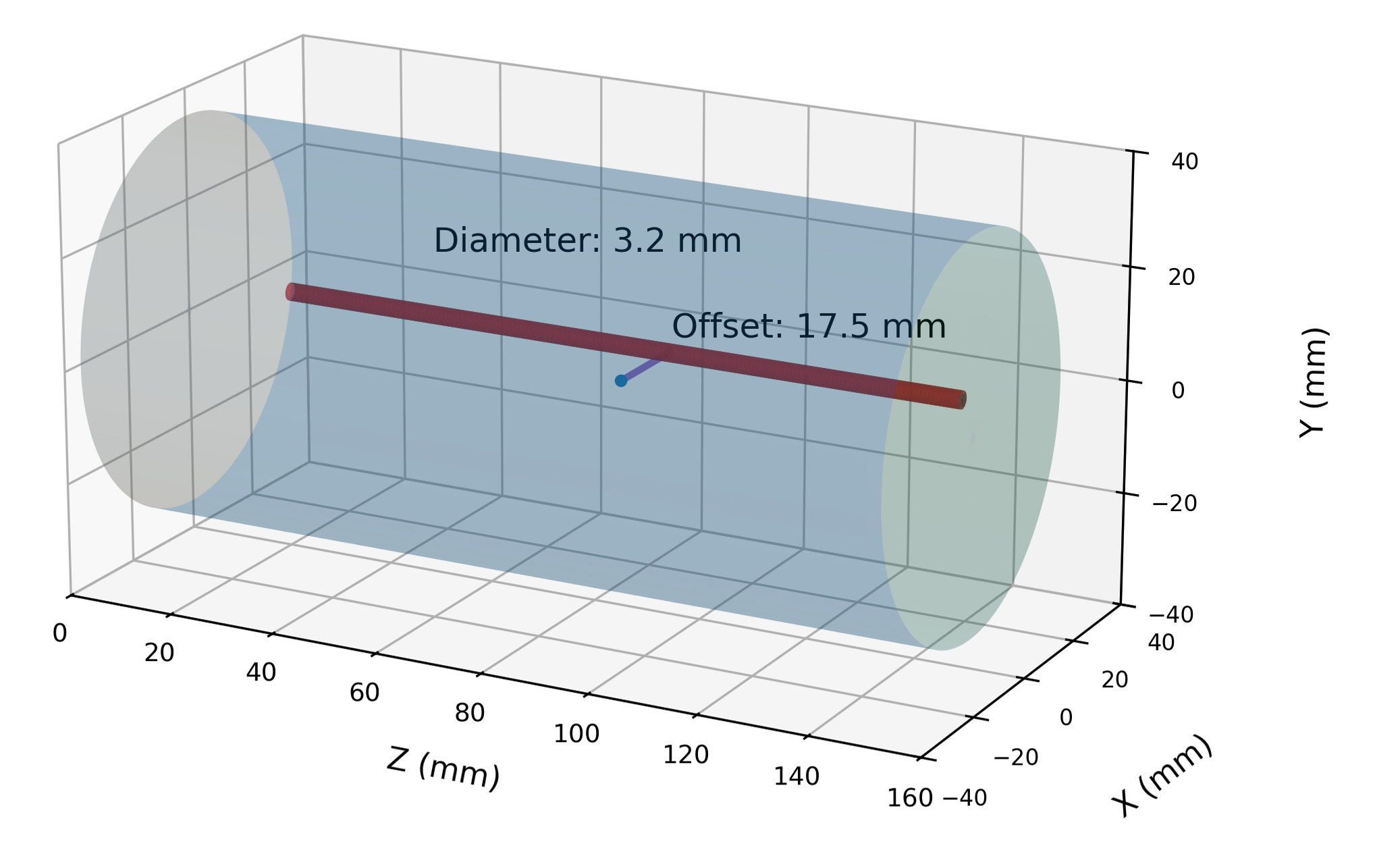}%
\label{fig4a}}
\hfil
\subfloat[]{\includegraphics[width=0.5\textwidth]{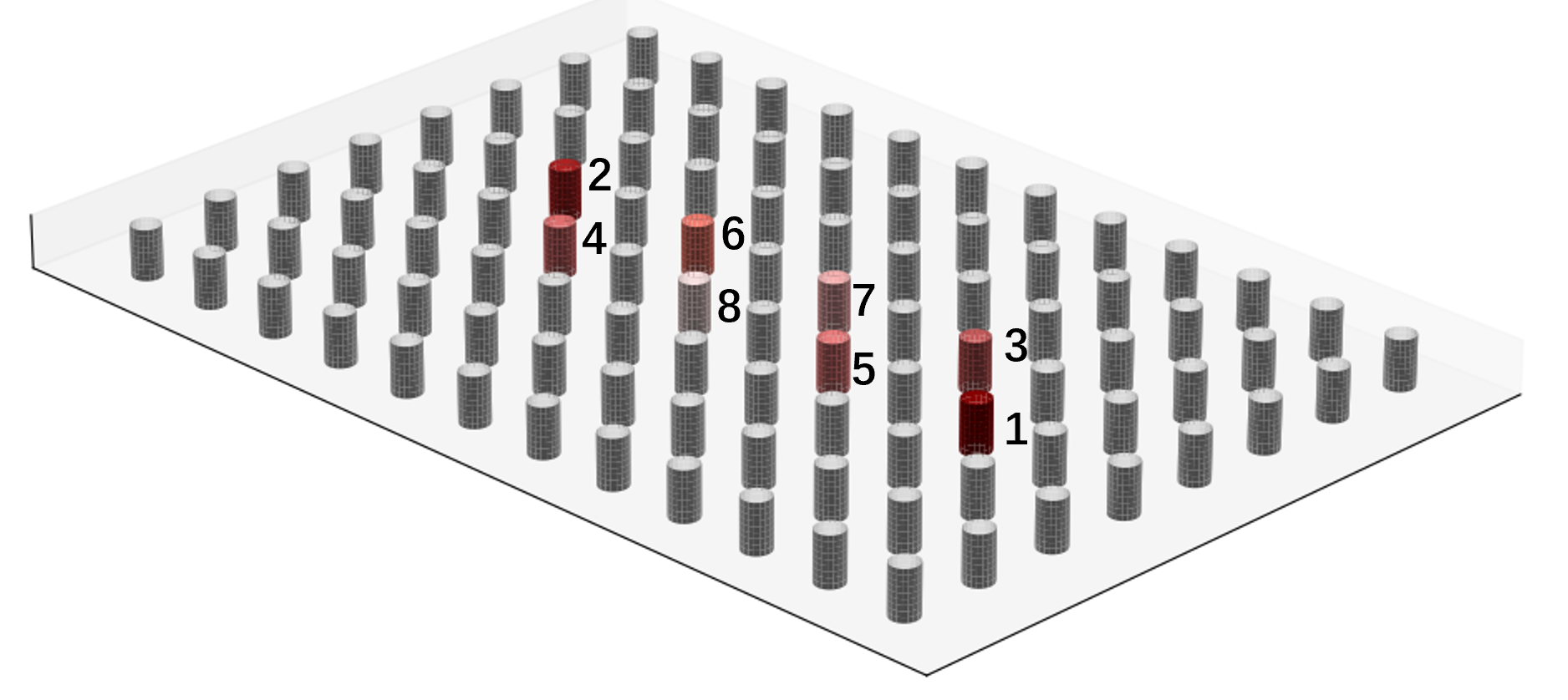}%
\label{fig4b}}
\label{fig4}
\caption{Phantoms used to calculate count rate and assess quantitative accuracy. (a) Structure of the polyethylene scatter phantom for count-rate measurement. (b) Radionuclide distribution in the 96-well plate phantom.}
\end{figure}

Scatter fraction and count-rate performance were evaluated using a cylindrical polyethylene scatter phantom with a diameter of $70 \pm 0.5$ mm and a length of $150 \pm 0.5$ mm. A cylindrical hole, 3.2 mm in diameter, was drilled parallel to the longitudinal axis at a radial offset of 17.5 mm from the central axis. The 70-mm diameter was selected based on our previous work~\cite{24} and differed from that specified for the NEMA NU 4-2008 rat phantom.
Background radiation was first measured for 30 minutes without a radioactive source. Subsequently, a line source filled with $^{68}$Ga solution, with an initial activity of 72.8 MBq (1.97 $m$Ci), was inserted into the off-center hole to simulate realistic scattering conditions, as shown in Fig.\ref{fig4a}. Data were acquired for 5 min at 10-min intervals until the measured count rate decreased to below twice the background level. A total of 60 acquisition frames were recorded.

Based on the acquisition with and without the radioactive source, sinograms were constructed via SSRB and used to extract the following metrics: total prompt count rate $R^{TOT}$, scatter count rate $R^s$, true count rate $R^t$, random count rate $R^r$, and noise equivalent count rate (NECR). According to the NEMA NU 4-2008, the calculation formula for the random count rate in double-coincidence measurements was as follows:
\begin{equation}
    \label{eqn_8}
    \left\{
    \begin{aligned}
        R_{i,j}^t = \frac{C_{i,j}^{TOT}-C_{i,j}^{r+s}}{T_{acq,j}}\\
        R_{i,j}^r = R_{i,j}^{TOT}-\frac{R_{i,j}^t}{1-SF_i}
    \end{aligned}
    \right.
\end{equation}
where $i$ and $j$ represent the slice index and acquisition frame, respectively. $C_{i,j}^{TOT}$ denotes the total count of coincidence events in slice $i$ of acquisition $j$. $C_{i,j}^{r+s}$ is the sum of random and scattered events estimated from the sinogram tails. 
The scatter fraction (SF) was calculated as
\begin{equation}
\label{eqn8}
    SF_i=\frac{\sum_{j^{'}}C_{i,j^{'}}^{r+s}}{\sum_{j^{'}}C_{i,j^{'}}^{TOT}}
    \approx\frac{\sum_{j^{'}}C_{i,j^{'}}^{s}}{\sum_{j^{'}}C_{i,j^{'}}^{t+s}}
\end{equation}
where $j^{'}$ is the final low-activity frames, for which random events were assumed to be negligible. In this study, the last five acquisitions were used for this calculation.

For triple-coincidence measurements, the scatter fraction could not be reliably estimated directly because of the limited event statistics. Since the LORs of triple-coincidence events are still defined by the two 511-keV annihilation photons, the scatter fraction measured from double-coincidence data, $SF_i$, was adopted as an approximation for the annihilation-photon pair in triple coincidences. However, the random count-rate estimation differs between double and triple coincidences because triple-coincidence data are additionally affected by CRTs. The consistent delayed-window method was used to estimate the random-triple count rate, primarily accounting for CRTs. This estimate was used in the subsequent count-rate analysis under the assumption that CRTs dominate the random-triple background. The count-rate components for triple coincidences were then calculated as
\begin{equation}
\label{eqn9}
    \left\{
    \begin{array}{l}
        R_{\mathrm{T},i,j}^s=SF_i(R_{\mathrm{T},i,j}^{TOT}-R_{\mathrm{T},i,j}^{r}- R_{\mathrm{T},i}^{\mathrm{int}})\\\\
        R_{\mathrm{T},i,j}^t=(\frac{1}{SF_i}-1)R_{\mathrm{T},i,j}^s
    \end{array}
    \right.
\end{equation}
where $R_{\mathrm{T},i,j}^{TOT}$ is the total triple-coincidence count rate, $R_{\mathrm{T},i,j}^{r}$ is the random-triple count rate estimated using the consistent delayed-window method, and $R_{\mathrm{T},i}^{\mathrm{int}}$ denotes the net intrinsic background contribution. The latter was calculated from the 30-min source-free acquisition as the prompt triple-coincidence count rate minus the corresponding random-triple count rate estimated using the same consistent delayed-window method.
From these measurements, count rate curves were derived for both double and triple coincidence modes.

\subsubsection{Quantitative Accuracy}
Quantitative accuracy was evaluated by assessing the linear relationship between reconstructed image intensity and true radionuclide activity. A 96-well plate phantom was utilized, following the methodology described in\cite{25}. Eight cylindrical wells within the plate were filled with $^{68}$Ga solutions of decreasing activity levels. Specifically, the activities from well 1 to well 8 were 60 $\mu$ Ci, 30 $\mu$Ci, 15 $\mu$Ci, 7 $\mu$Ci, 3.5 $\mu$Ci, 1.75 $\mu$Ci, 0.875 $\mu$Ci, and 0.437 $\mu$Ci, respectively, as shown in Fig. \ref{fig4b}. The phantom was positioned at the center of the system FOV and scanned in three consecutive 10-minute acquisitions. Radioactive decay correction for $^{68}$Ga was applied to all three datasets before quantitative analysis.

Following image reconstruction as mentioned in Section 2.3, both double- and triple-coincidence datasets were analyzed. Regions of interest (ROIs) were drawn over each of the eight wells, and the mean voxel value within each ROI was plotted against the known activity to assess quantitative linearity for both double- and triple-coincidence reconstructions.

\subsubsection{Dual-Isotope Separation Performance}
Dual-isotope separation performance was evaluated to determine whether the proposed system could recover and distinguish the spatial distributions of two radionuclides from a single acquisition. A background-free acrylic phantom was employed for this experiment. The phantom contained nine rectangular cavities spaced at 10-mm intervals within a 34 mm-long block. Four cavities were filled with radionuclides: two with $^{68}$Ga solution, with activities of 90 $\mu$Ci and 102 $\mu$Ci, and two with $^{18}$F solution, with activities of 38 $\mu$Ci and 20 $\mu$Ci.

A 15-minute acquisition was performed, and images were reconstructed using 20 iterations of the MLEM/CDEM algorithm. Image-domain dual-isotope separation was then performed using the scaling-factor-based method described in Section 2.4. Separation performance was quantitatively evaluated using the signal preservation ratio and residual signal ratio.
The signal preservation ratio was defined as the fraction of signal retained in the correct separated image, whereas the residual signal ratio was defined as the fraction of cross-contaminating signal remaining in the incorrect separated image. Specifically, for $^{18}$F-filled cavities, the signal preservation ratio was calculated as the ratio of the signal intensity in the $^{18}$F-separated image to that in the conventional double-coincidence image, whereas the residual signal ratio was calculated as the ratio of the residual signal intensity in the $^{68}$Ga-separated image to that in the conventional double-coincidence image. The preservation and residual signal ratios for $^{68}$Ga-filled cavities were calculated in the same manner. For dual-isotope images separated in image domain, the sum of the preservation and residual signal ratios is expected to be 100\%.

\subsubsection{In Vivo Animal Study}
To evaluate the \textit{in vivo} imaging performance of the proposed dual-isotope PET system, two sets of small-animal imaging experiments were conducted using $^{18}$F/$^{124}$I and $^{18}$F/$^{68}$Ga dual-tracer combinations. For the animal experiments, the signal preservation and residual signal ratios were interpreted as relative signal fractions within tracer-dominant organ ROIs, with the single-tracer images serving as qualitative references for the expected uptake patterns.
All animal procedures were approved by the Institutional Animal Care and Use Committee of Air Force Medical University (Fourth Military Medical University) with approval No. 251623.

\paragraph{$^{18}$F-FDG/$^{124}$I Dual-Isotope Imaging}
In the first experiment, a healthy mouse (Mouse A) was intravenously injected with 100 $\mu$Ci (3.7 MBq) of $^{18}$F-FDG. After a 1-hour uptake period under anesthesia, a 15-minute scan was performed to obtain the reference $^{18}$F-FDG image. Another healthy mouse (Mouse B) was injected with 100 $\mu$Ci (3.7 MBq) of $^{124}$I, followed by the same imaging protocol to acquire the $^{124}$I reference image. Subsequently, Mouse B was injected with an additional 100 $\mu$Ci (3.7 MBq) of $^{18}$F-FDG. A 15-minute dual-isotope scan was performed after a 1-h uptake period. 

For the $^{18}$F-FDG/$^{124}$I dual-isotope acquisition, a 350--650-keV energy window was used to select annihilation photons, whereas a 500--800-keV energy window was applied to select prompt gamma candidates from $^{124}$I. Given the overlap between these energy windows, geometric constraints based on our previously developed method~\cite{Shi2022PromptSpectrum} were employed to identify annihilation-photon pairs and construct the corresponding LORs. The double- and triple-coincidence data were then reconstructed using 20 iterations of MLEM and CDEM, respectively, followed by image-domain isotope separation.

\paragraph{$^{18}$F-FDG/$^{68}$Ga-PSMA Dual-Isotope Imaging}
Three healthy mice were used in this group. The first mouse received an intravenous injection of 330 $\mu$Ci (12.2 MBq) of $^{68}$Ga-PSMA and underwent a 30-minute scan after injection to obtain the reference $^{68}$Ga image. The second mouse received 126 $\mu$Ci (4.7 MBq) of $^{18}$F-FDG and was scanned using the same procedure to acquire the reference $^{18}$F image. The third mouse was simultaneously injected with 316 $\mu$Ci (11.7 MBq) of $^{68}$Ga-PSMA and 72 $\mu$Ci (2.7 MBq) of $^{18}$F-FDG. A 30-min dual-isotope scan was performed after injection. Data processing, reconstruction and separation procedures followed the same protocol as described above.

\section{Results}
\subsection{Energy and Timing Resolution}
The system achieved energy resolutions of 15.7\% at 511 keV, 12.9\% at 1077 keV and 10.2\% at 1275 keV, respectively, as measured using $^{68}$Ga and $^{22}$Na point sources. The coincidence timing resolution was determined from the time-difference histogram of coincidence events. A Gaussian function was fitted to the histogram, and the FWHM of the fitted curve was calculated to be 1.35 ns independent of the radionuclide used.

\subsection{Spatial Resolution}
Spatial resolution was evaluated using reconstructed images of a $^{22}$Na point source placed at multiple positions within the FOV. The FWHM and FWTM of the point spread function were measured along the $X$, $Y$, and $Z$ directions from line profiles drawn through the peak signal intensity.

\begin{figure}[htbp]
\centering
\subfloat[]{\includegraphics[width=1.0\textwidth]{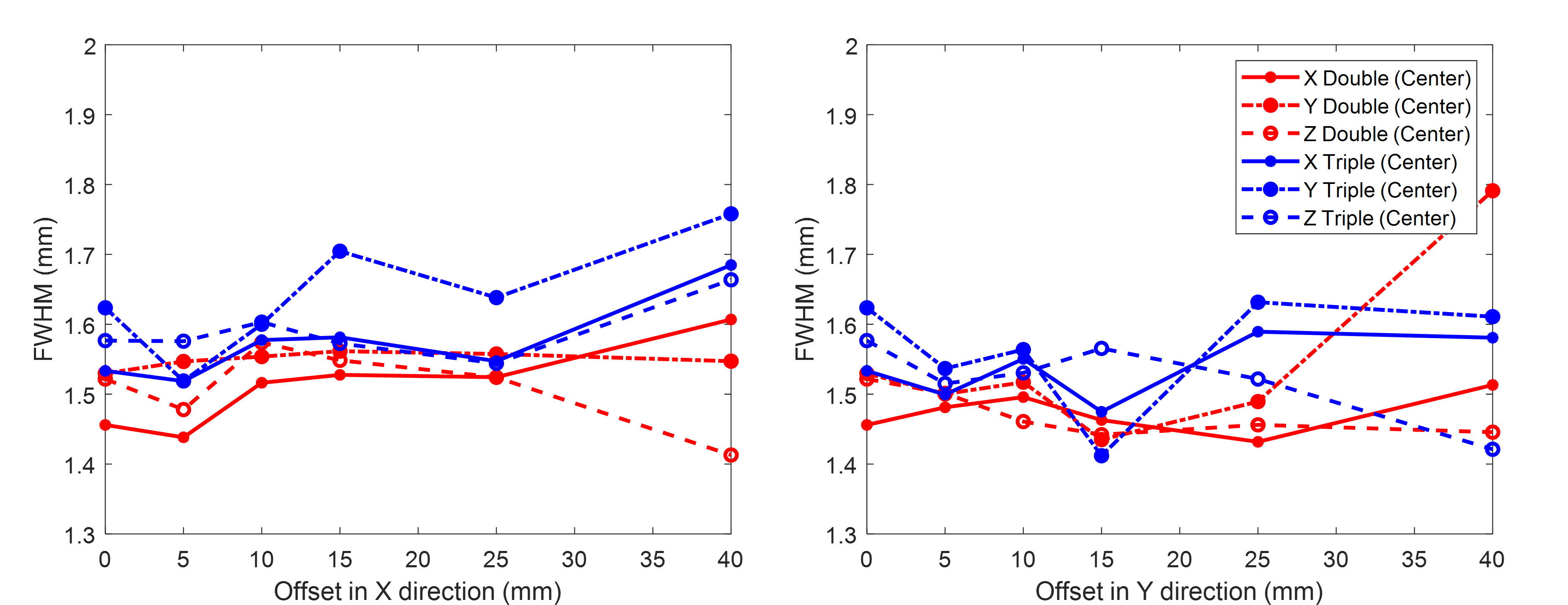}%
\label{fig5a}}
\hfil
\subfloat[]{\includegraphics[width=1.0\textwidth]{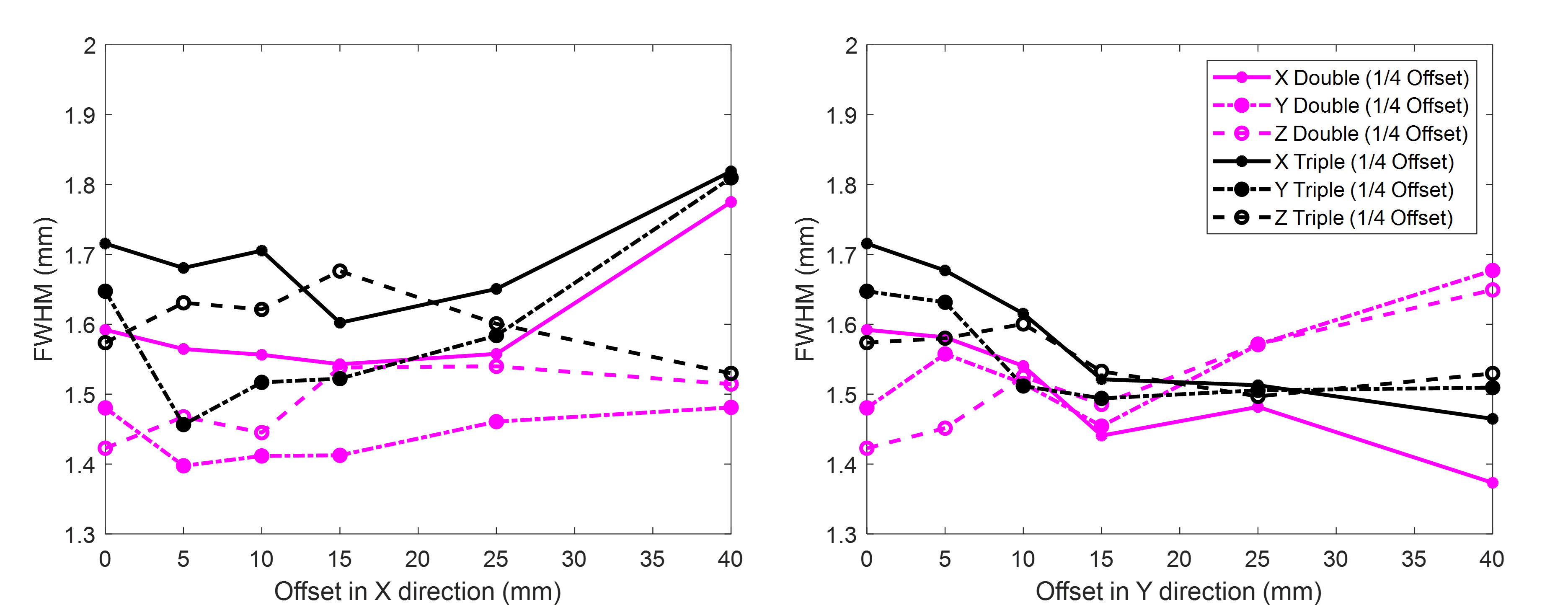}%
\label{fig5b}}
\caption{Spatial resolution in the X, Y, and Z directions as a function of the offset along the X/Y axis. (a) FWHM of the point sources reconstructed using double or triple coincidences at the axial center of the FOV. (b) FWHM of the point sources reconstructed using double or triple coincidences at one-fourth of the axial FOV from the center.}
\label{fig5}
\end{figure}

As illustrated in Fig. \ref{fig5}, at the axial center of the FOV, the mean FWHM$\slash$FWTM values of the double-coincidence images were 1.49$\slash$2.72 mm, 1.55$\slash$2.82 mm, and 1.49$\slash$2.71 mm in the $X$, $Y$, and $Z$ directions, respectively. For triple-coincidence images, the corresponding spatial resolutions were 1.56$\slash$2.84 mm, 1.60$\slash$2.91 mm, and 1.55$\slash$2.84 mm, respectively. At one-fourth of the axial FOV from the center, the mean FWHM$\slash$FWTM values of double-coincidence images were 1.55$\slash$2.82 mm, 1.49$\slash$2.72 mm, and 1.50$\slash$2.74 mm in the $X$, $Y$, and $Z$ directions, respectively. The corresponding values for the triple-coincidence images were 1.64$\slash$2.99 mm, 1.57$\slash$2.86 mm, 1.58$\slash$2.88 mm, respectively.

\begin{figure}[htbp]
\centering
\subfloat[]{\includegraphics[width=0.5\textwidth]{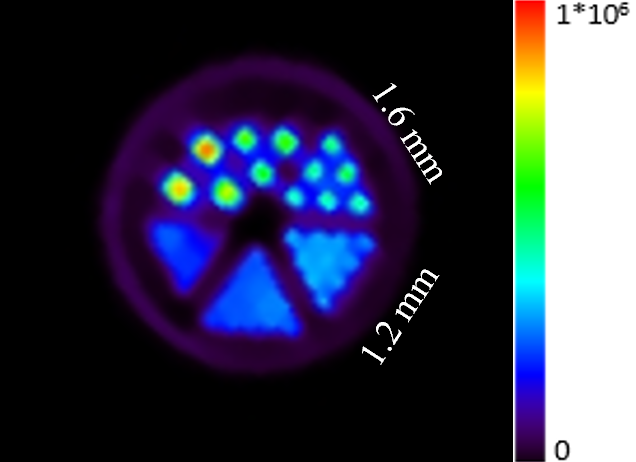}%
\label{fig6a}}
\subfloat[]{\includegraphics[width=0.5\textwidth]{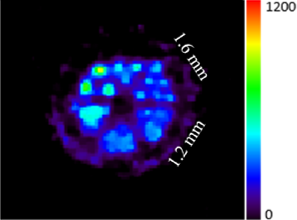}%
\label{fig6b}}
\caption{Derenzo phantom images reconstructed from (a) double-coincidence data and (b) triple-coincidence data.}
\label{fig6}
\end{figure}

To further validate the spatial resolution, a small Derenzo phantom was imaged using both coincidence modes. As shown in Fig. \ref{fig6}, the 1.6-mm rods were clearly resolved in the double-coincidence image, whereas the 1.2-mm rods were only partially distinguishable. In the triple-coincidence image, rods with a diameter of 1.6 mm remained discernible, although with reduced image contrast and increased noise. These findings confirm the high-resolution imaging capability of the proposed system in both coincidence modes, while also showing the expected resolution degradation in triple-coincidence imaging caused by limited triple-event statistics.

\subsection{Prompt Gamma Detection Efficiency}
\begin{figure}[htbp]
\centering
\includegraphics[width=0.9\textwidth]{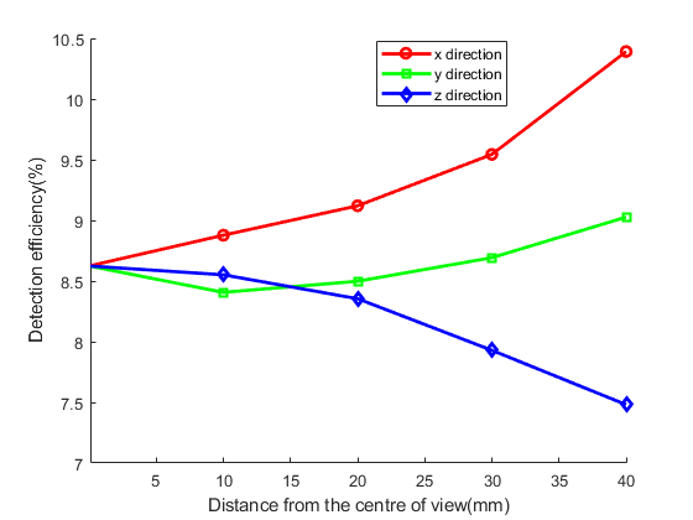}
\caption{Detection efficiency of prompt gamma using a $^{22}$Na point source (1275 keV).}
\label{fig_7}
\end{figure}

At the center of the FOV, a 10-minute acquisition of a $^{22}$Na point source captured approximately 1.6 $\times10^7$ double-coincidence events and 1.4 $\times 10^6$ triple-coincidence events. As shown in Fig. \ref{fig_7}, the average prompt gamma detection efficiency along the $X$ axis was 9.31\%, with a maximum value of 10.39\%. Along the $Y$ axis, the average and maximum efficiencies were 8.65\% and 9.03\%, respectively. Along the $Z$ axis, the average and maximum efficiencies were 8.19\% and 8.62\%, respectively. These results demonstrate that the system maintains effective detection of prompt gammas across the sampled positions, with only modest spatial variation.

\subsection{System Sensitivity}
\begin{figure}[htbp]
\centering
\subfloat[]{\includegraphics[width=0.75\textwidth]{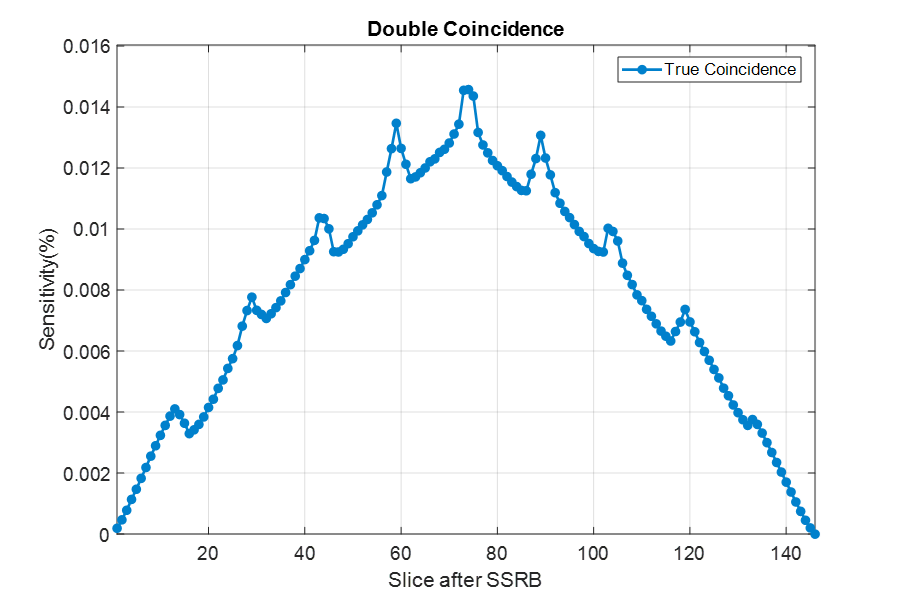}%
\label{fig8a}}\\
\subfloat[]{\includegraphics[width=0.75\textwidth]{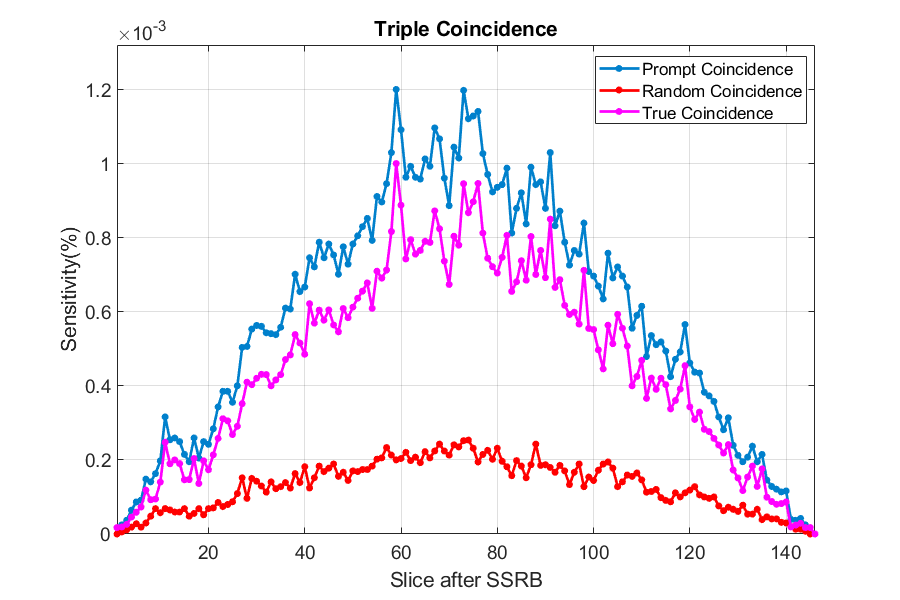}%
\label{fig8b}}
\caption{Axial sensitivity profile for double and triple coincidences. (a) Axial sensitivity profile for double coincidences with an energy window of 350-650 keV. (b) Axial sensitivity profile for triple coincidences with 350-650-keV energy window for annihilation gamma and 900--1400-keV energy window for prompt gamma.}
\label{fig8}
\end{figure}

System sensitivity was calculated as the true coincidence count rate detected per unit source activity, in accordance with the YY/T 1835-2022 standard. Fig. \ref{fig8} illustrates the axial sensitivity profiles across slices after SSRB. By summing the sensitivity values over all axial slices, the overall double-coincidence sensitivity was calculated to be 1.12\%. For triple-coincidence, the branching-normalized triple-coincidence sensitivity, calculated using the joint positron--gamma branching fraction of 1.2\% in Eq.~\ref{eqn_7}, was 0.085\% before random-triple correction. According to the standard protocol, only events located within 1.5 cm of the line-source center were included
in the sensitivity analysis. However, the existence of correlated random events makes the triple coincidences extracted by the above procedure still contain a considerable number of random-triple events. Therefore, the consistent delayed-window method was further applied to estimate and correct for random-triple events. 

As shwon in Fig.\ref{fig8b}, after CRT correction, the sensitivity of true coincidences decreased to 0.068\%. Based on the ratio between the corrected triple-coincidence sensitivity and the double-coincidence sensitivity, the average detection efficiency of the prompt gammas along the $Z$ axis was estimated to be 6.07\%. Limiting the axial FOV to the central 80 mm increased the average efficiency slightly to 6.34\%, which was lower than that measured in the previous $^{22}$Na scanning. 
This discrepancy should be interpreted cautiously because the two values were obtained using different measurement protocols. The $^{22}$Na result represents a direct point-source estimate of prompt gamma detection efficiency, whereas the $^{68}$Ga result was inferred from the ratio of triple- to double-coincidence sensitivities measured with a line source. Therefore, the lower $^{68}$Ga value more likely reflects protocol-dependent effects, including source geometry, energy-window acceptance, and ROI-based event selection, rather than a reduction in the intrinsic prompt gamma detection capability of the detector.

\subsection{Scatter Fraction and Count Rate Performance}
Based on the NEMA NU 4-2008 calculation procedure, SF for double-coincidence events was determined to be 27.81\% using a polyethylene scatter phantom containing a $^{68}$Ga line source. Due to the limited statistics of triple-coincidence events, the SF for triple coincidence was not directly measured. Since the LORs of triple-coincidence events are still defined by the two 511-keV annihilation photons, the scatter fraction measured from double-coincidence data was adopted as an approximation for the annihilation-photon pair in subsequent triple-coincidence count rate analysis.

\begin{figure}[!htbp]
\centering
\subfloat[]{\includegraphics[width=0.5\textwidth]{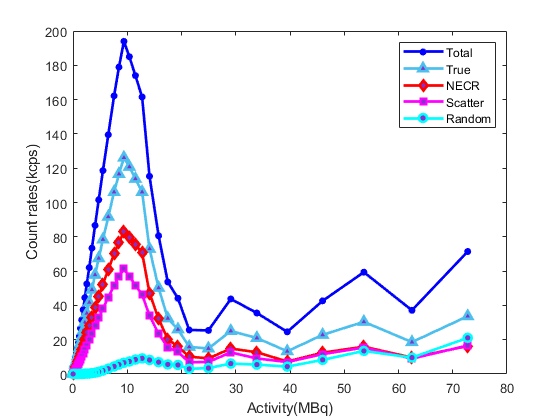}%
\label{fig9a}}\\
\subfloat[]{\includegraphics[width=0.5\textwidth]{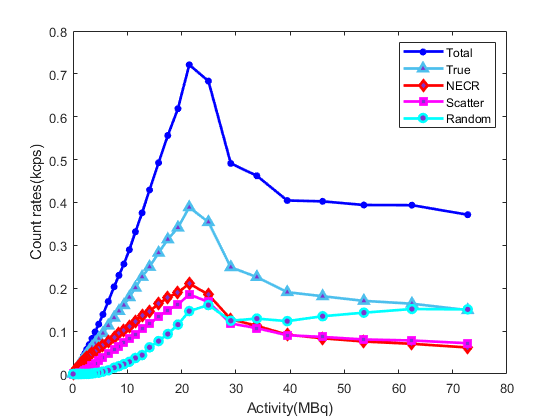}%
\label{fig9b}}
\subfloat[]{\includegraphics[width=0.5\textwidth]{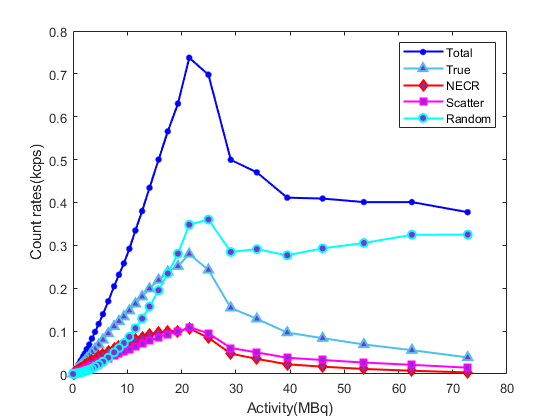}%
\label{fig9c}}
\caption{Measured total, true, random, scatter and noise-equivalent count rates as functions of activity using a polyethylene scatter phantom containing a $^{68}$Ga line source. (a) Count rate for double coincidences where random count rate was estimated using the NEMA NU4-2008 procedure. (b) Count rate for triple coincidences where random count rate was estimated using the NEMA-style procedure. (c) Count rate for triple coincidences where random count rate was estimated by consistent delayed window.}
\label{fig9}
\end{figure}

For double coincidences, the total count rate and NECR increased with source activity, reaching peak values of 194 kcps and 83 kcps, respectively, at 9.38 MBq, as shown in Fig.~\ref{fig9a}. Beyond this activity, both count rates decreased sharply but exhibited nonmonotonic variations at higher activities.
For triple coincidences, when the NEMA-based random estimation procedure was applied, the total count rate and NECR reached maximum values of 721 cps and 211 cps, respectively, at an activity of 21.42 MBq, as shown in Fig.\ref{fig9b}.

However, when the effect of CRTs was considered, the random count rate increased rapidly with activity and eventually stabilized at approximately 325 cps, as shown in Fig.\ref{fig9c}. At an activity of 17.47 MBq, the random count rate was nearly equal to the true count rate, with a value of 237 cps. As the activity increased further, random events dominated the total counts, accounting for over 85\% at the highest tested activity of 72.8 MBq. After correction using this procedure, the peak NECR for triple coincidences was 106 cps at 21.42 MBq.

\subsection{Quantitative Accuracy}
Quantitative accuracy was evaluated by assessing the linear relationship between the reconstructed voxel intensity and the true radionuclide activity. A 96-well plate phantom containing eight wells filled with decreasing activity of $^{68}$Ga solution, ranging from 60 $\mu$Ci to 0.437 $\mu$Ci, was scanned over three consecutive 10-minute acquisitions.

\begin{figure}[!htbp]
\centering
\subfloat[]{\includegraphics[width=1.0\textwidth]{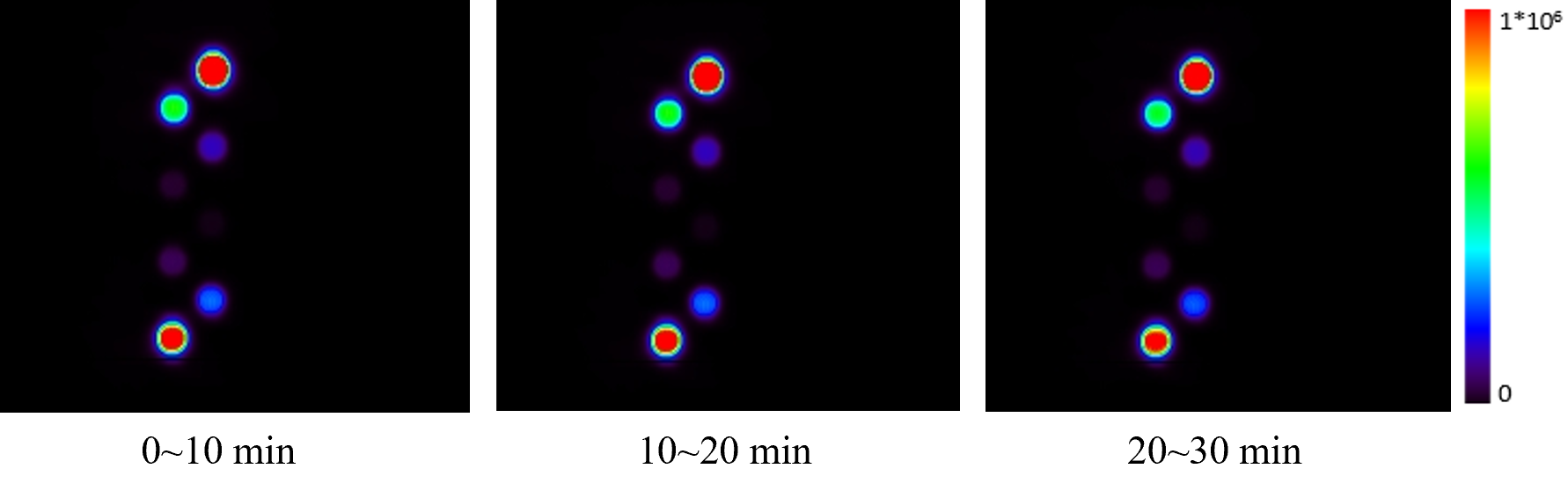}%
\label{fig10a}}
\hfil
\subfloat[]{\includegraphics[width=1.0\textwidth]{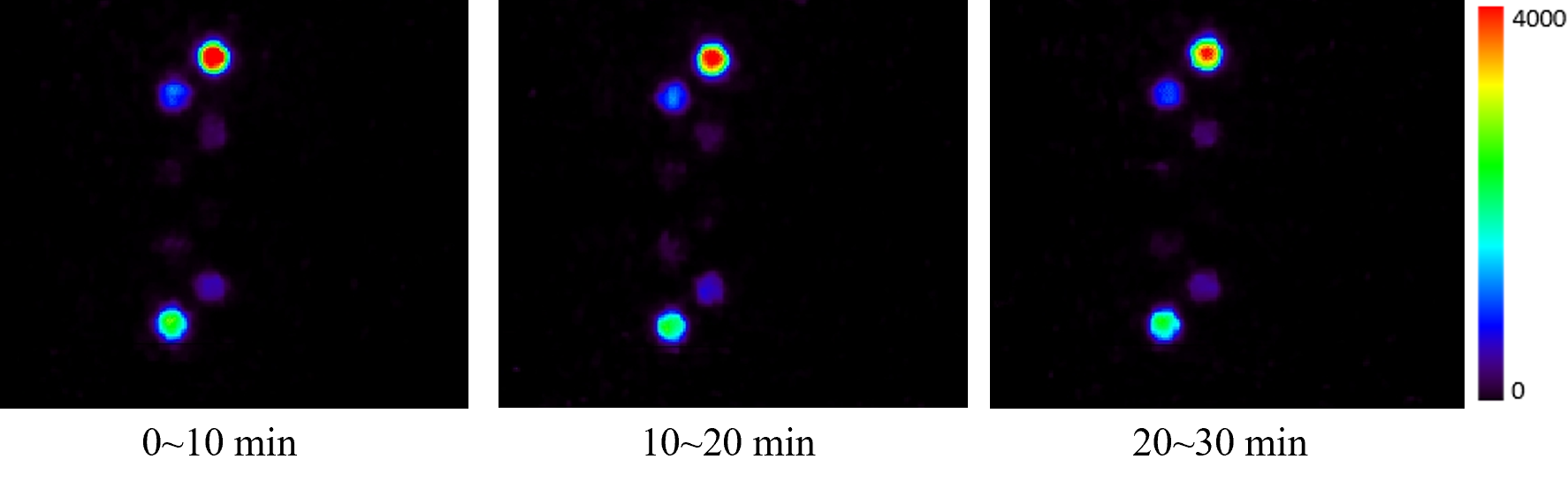}%
\label{fig10b}}
\hfil
\subfloat[]{\includegraphics[width=0.5\textwidth]{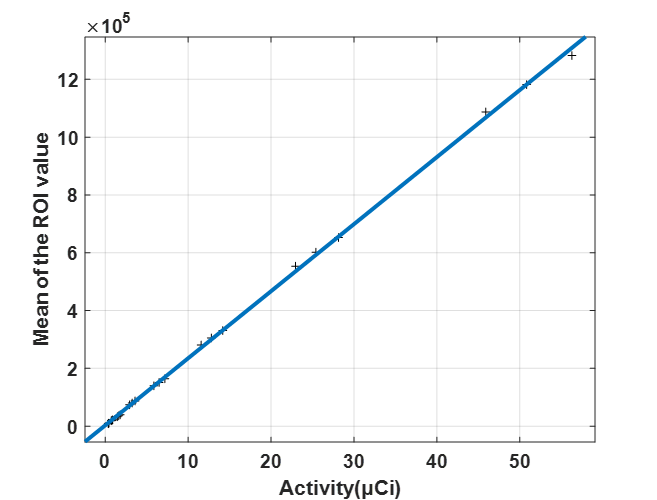}%
\label{fig10c}}
\subfloat[]{\includegraphics[width=0.5\textwidth]{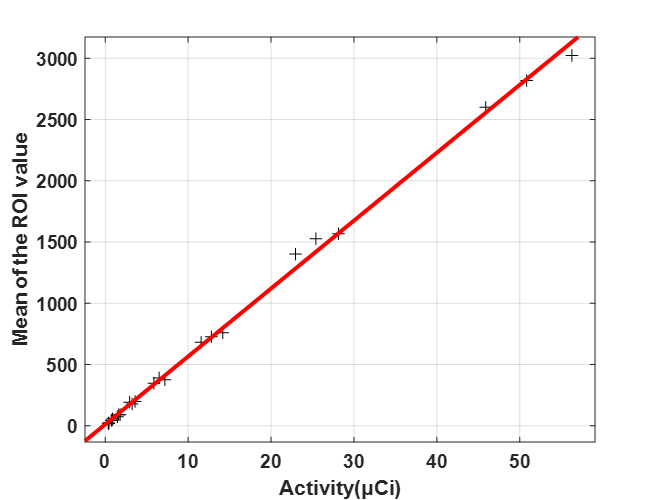}%
\label{fig10d}}
\caption{Quantitative accuracy assessment using a 96-well plate phantom filled with $^{68}$Ga. (a) Reconstructed double-coincidence images acquired during the 0--10 minute, 10--20 minute and 20--30 minute intervals. (b) Reconstructed triple-coincidence images acquired during the same intervals. (c) Linear fitting of mean ROI values from double-coincidence images versus true activities. (d) Linear fitting of mean ROI values from triple-coincidence images versus true activities.}
\label{fig10}
\end{figure}

Fig.\ref{fig10a} displays the reconstructed double-coincidence images for the 0–-10 minute, 10–-20 minute, and 20–-30 minute intervals, respectively. ROIs were drawn over the eight wells at different time periods, then the mean voxel values within each ROI were extracted and plotted against the corresponding true activity. The $X$ axis represents the known activity in $\mu$Ci and the $Y$ axis denotes the mean voxel value in the corresponding ROI. Linear regression analysis showed excellent linearity for double-coincidence reconstruction, with an $R^2$ value of 0.9995, as shown in Fig.\ref{fig10c}.

Similarly, Fig.\ref{fig10b} presents the triple-coincidence images reconstructed from the same time intervals. Despite the substantially lower count statistics of triple coincidences, the mean ROI values remained highly correlated with the true activities. Linear fitting produced an $R^2$ of 0.9978 as shown in Fig\ref{fig10d}. 
These results demonstrate strong linearity between reconstructed ROI intensity and source activity for both double- and triple-coincidence imaging over the investigated activity range.

\subsection{Dual-Isotope Separation}
\subsubsection{Phantom Experiment}
Quantitative analysis of the separation performance is summarized in Table \ref{tab1}(a). In the $^{68}$Ga-separated image, the signal preservation ratios for the two $^{68}$Ga-filled wells were 93.58\% and 97.97\%, while the residual signal ratios at the two $^{18}$F-filled wells were 1.54\% and 2.05\%, respectively.
In the $^{18}$F-separated image, the signal preservation ratios for the two $^{18}$F-filled wells were 98.46\% and 97.95\%, with corresponding residual signal ratios of 6.42\% and 2.03\% at the $^{68}$Ga-filled wells. 
The signal preservation ratios for both radionuclides exceeded 90\%, demonstrating that the proposed system can separate $^{18}$F and $^{68}$Ga signals with high fidelity at the level of physical phantom.
The slightly lower signal preservation ratio for $^{68}$Ga may be attributed to the limited statistics of triple-coincidence events, indicating that further optimization of low-count triple-coincidence reconstruction may improve separation performance.
\begin{table}[htbp]
\caption{Signal preservation and residual signal ratios (\%) in phantom and mouse experiments. Blue-shaded entries indicate signal preservation ratios, while the remaining entries indicate residual signal ratios. For the mouse experiments, the ratios represent relative signal allocation within tracer-dominant organ ROIs.}
\label{tab1}
\centering
\renewcommand{\arraystretch}{1.2}
\setlength{\tabcolsep}{2pt}
\small

\begin{tabular}{|C{2.7cm}|C{1.4cm}|C{1.4cm}|C{1.4cm}|C{1.4cm}|}
\multicolumn{5}{c}{\textbf{(a) $^{68}$Ga/$^{18}$F phantom experiment}}\\
\hline
\diagbox[
width=2.7cm,
height=1.0cm
]{\hspace{-0.5em}Reconstruction}{Isotope}
& \multicolumn{2}{c|}{$^{68}$Ga isotope}
& \multicolumn{2}{c|}{$^{18}$F isotope} \\
\hhline{|~|----|}
& 1 & 2 & 1 & 2 \\
\hline
$^{68}$Ga image
& \cellcolor{blue!30}93.58
& \cellcolor{blue!30}97.97
& 1.54
& 2.05 \\
\hline
$^{18}$F image
& 6.42
& 2.03
& \cellcolor{blue!30}98.46
& \cellcolor{blue!30}97.95 \\
\hline
\end{tabular}

\vspace{3mm}

\begin{tabular}{|C{2.7cm}|C{2.93cm}|C{2.93cm}|}
\multicolumn{3}{c}{\textbf{(b) $^{124}$I/$^{18}$F mouse experiment}}\\
\hline
\diagbox[
width=2.7cm,
height=1.0cm,
trim=l,
innerleftsep=1pt,
innerrightsep=1pt
]{\hspace{-0.5em}{Reconstruction}}{\hspace{-0.0em}Isotope}
& \makecell{$^{124}$I isotope \\ (Thyroid)}
& \makecell{$^{18}$F isotope \\ (Heart)} \\
\hline
$^{124}$I image
& \cellcolor{blue!30}85.00
& 19.55 \\
\hline
$^{18}$F image
& 15.00
& \cellcolor{blue!30}80.45 \\
\hline
\end{tabular}

\vspace{3mm}

\begin{tabular}{|C{2.7cm}|C{2.93cm}|C{2.93cm}|}
\multicolumn{3}{c}{\textbf{(c) $^{68}$Ga/$^{18}$F mouse experiment}}\\
\hline
\diagbox[
width=2.7cm,
height=1.0cm,
trim=l,
innerleftsep=1pt,
innerrightsep=1pt
]{\hspace{-0.5em}{Reconstruction}}{\hspace{-0.0em}Isotope}
& \makecell{$^{68}$Ga isotope \\ (Kidney)}
& \makecell{$^{18}$F isotope \\ (BAT)} \\
\hline
$^{68}$Ga image
& \cellcolor{blue!30}77.78
& 5.76 \\
\hline
$^{18}$F image
& 22.22
& \cellcolor{blue!30}94.24 \\
\hline
\end{tabular}
\end{table}

\subsubsection{In Vivo Animal Study}
\begin{figure}[!htbp]
\centering
\subfloat[]{\includegraphics[height=1.8in]{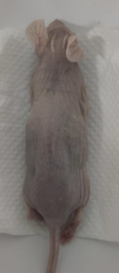}%
\hspace{1mm}\label{fig11a}}
\subfloat[]{\includegraphics[height=1.8in]{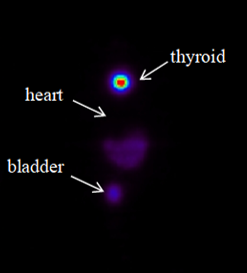}%
\hspace{1mm}\label{fig11b}}
\subfloat[]{\includegraphics[height=1.8in]{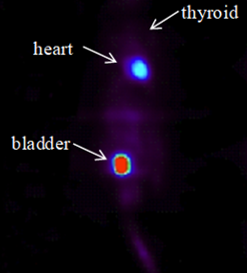}%
\hspace{1mm}\label{fig11c}}\\
\subfloat[]{\includegraphics[height=1.8in]{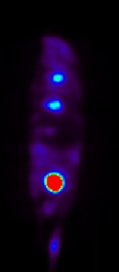}%
\hspace{1mm}\label{fig11d}}
\subfloat[]{\includegraphics[height=1.8in]{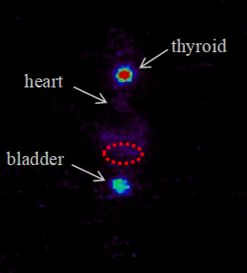}%
\hspace{1mm}\label{fig11e}}
\subfloat[]{\includegraphics[height=1.8in]{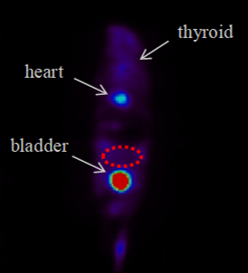}%
\hspace{1mm}\label{fig11f}}
\caption{$^{18}$F/$^{124}$I dual-isotope mouse imaging results. (a) Optical image of the mouse. (b) $^{124}$I single-isotope reference reconstructed from double coincidences. (c) $^{18}$F single-isotope reference reconstructed from double coincidences. (d) Conventional double-coincidence reconstruction of the $^{18}$F/$^{124}$I dual-isotope acquisition. (e) Separated $^{124}$I image. (f) Separated $^{18}$F image obtained by image-domain subtraction.}
\label{fig11}
\end{figure}

\paragraph{$^{18}$F-FDG/$^{124}$I dual-isotope imaging}
Fig. \ref{fig11} shows the results of the \textit{in vivo} $^{18}$F/$^{124}$I dual-isotope imaging in mice. In the conventional double-coincidence reconstruction of the dual-isotope acquisition, shown in Fig.~\ref{fig11d}, the signals from the two radionuclides were superimposed and could not be distinguished without prior anatomical or physiological information.
In contrast, after isotope separation, the $^{124}$I-separated image shown in Fig.~\ref{fig11e} exhibited a dominant signal in the thyroid region and suppressed signal in the heart region, consistent with the $^{124}$I single-isotope reference image shown in Fig.~\ref{fig11b}.

Similarly, the $^{18}$F-separated image shown in Fig.~\ref{fig11f} recovered the characteristic $^{18}$F-FDG uptake in the heart and bladder regions, consistent with the $^{18}$F single-isotope reference image shown in Fig.~\ref{fig11c}.
These results demonstrate that the proposed system can recover the characteristic uptake patterns of $^{124}$I and $^{18}$F-FDG from a simultaneous dual-isotope acquisition, supporting the feasibility of \textit{in vivo} tracer separation.

Quantitative analysis was performed by outlining ROIs over the thyroid and heart, which were treated as $^{124}$I- and $^{18}$F-FDG-dominant regions, respectively, with reference to the single-tracer images. As summarized in Table~\ref{tab1}(b), the $^{124}$I-separated image achieved an 85.00\% signal preservation ratio in the thyroid region, with a residual signal ratio of 19.55\% in the heart region. For the $^{18}$F-separated image, the heart signal was preserved at 80.45\%, with a residual signal ratio of 15.00\% in the thyroid region. These results confirm the feasibility of \textit{in vivo} separation of $^{124}$I and $^{18}$F using the proposed system.

\begin{figure}[!htbp]
\centering
\subfloat[]{\includegraphics[height=1.8in]{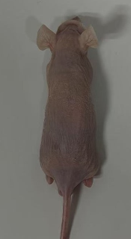}%
\hspace{1mm}\label{fig12a}}
\subfloat[]{\includegraphics[height=1.8in]{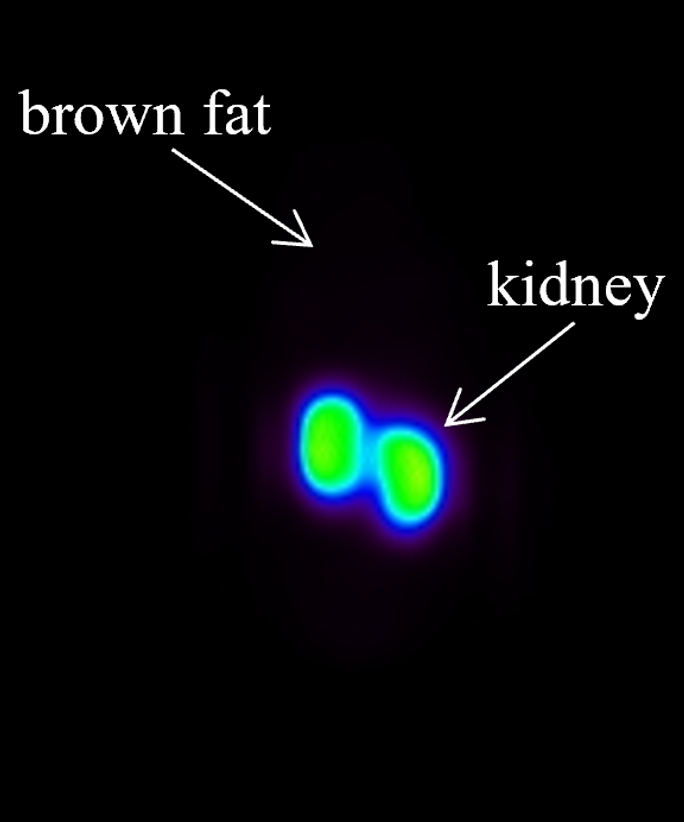}%
\hspace{1mm}\label{fig12b}}
\subfloat[]{\includegraphics[height=1.8in]{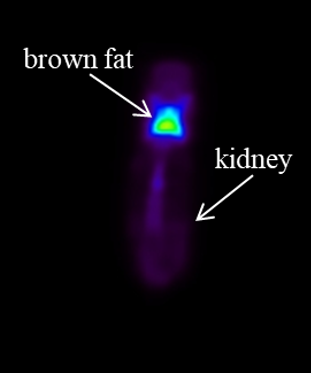}%
\hspace{1mm}\label{fig12c}}\\
\subfloat[]{\includegraphics[height=1.8in]{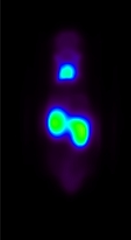}%
\hspace{1mm}\label{fig12d}}
\subfloat[]{\includegraphics[height=1.8in]{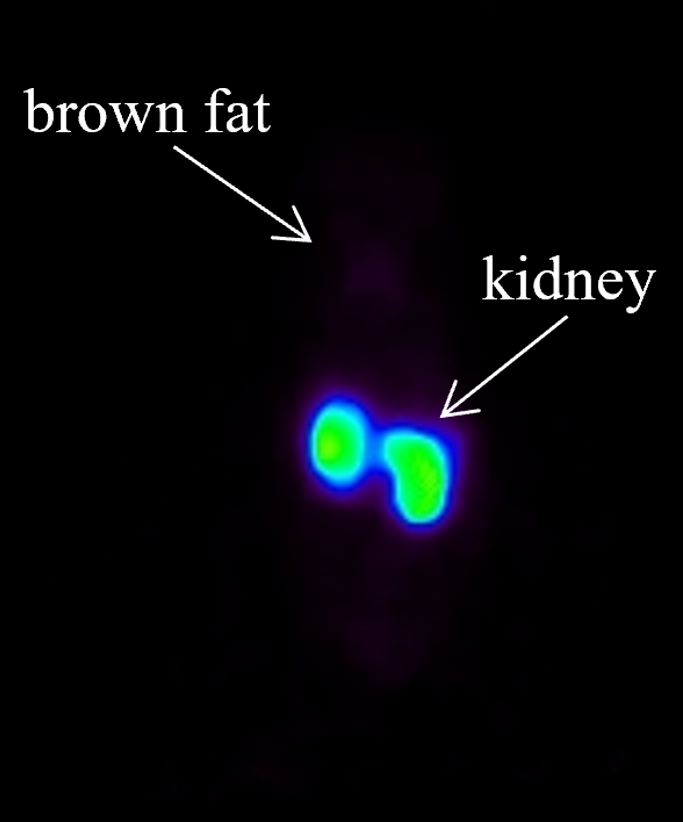}%
\hspace{1mm}\label{fig12e}}
\subfloat[]{\includegraphics[height=1.8in]{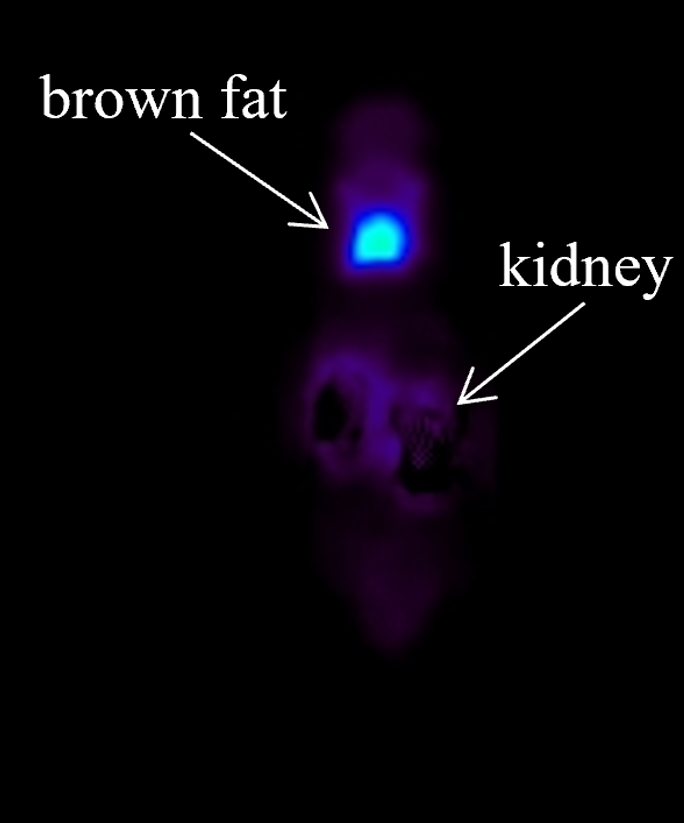}%
\hspace{1mm}\label{fig12f}}
\caption{$^{18}$F/$^{68}$Ga dual-isotope mouse imaging results. (a) Optical image of the mouse. (b) $^{68}$Ga single-isotope reference reconstructed from double coincidences. (c) $^{18}$F single-isotope reference reconstructed from double coincidences. (d) Conventional double-coincidence reconstruction of the $^{18}$F/$^{68}$Ga dual-isotope acquisition. (e) Separated $^{68}$Ga image. (f) Separated $^{18}$F image obtained by image-domain subtraction.}
\label{fig12}
\end{figure}

\paragraph{$^{18}$F-FDG/$^{68}$Ga-PSMA dual-isotope imaging} 
Fig. \ref{fig12} presents the \textit{in vivo} dual-isotope imaging results for $^{18}$F and $^{68}$Ga in mice. Figs. \ref{fig12b} and \ref{fig12c} display the single-isotope reference images for $^{68}$Ga-PSMA and $^{18}$F-FDG, respectively. As expected, $^{68}$Ga-PSMA shows predominant uptake in the kidneys, whereas $^{18}$F-FDG shows metabolic uptake, particularly in brown adipose tissue (BAT) in the displayed slice.

In the conventional double-coincidence reconstruction of the dual-isotope acquisition, shown in Fig. \ref{fig12d}, both kidney and BAT regions exhibited high uptake, making it difficult to differentiate the two tracers from the mixed image alone.
After isotope separation using triple coincidences, the $^{68}$Ga-separated image shown in Fig. \ref{fig12e} exhibited activity predominantly localized in the kidneys, whereas the $^{18}$F-separated image shown in Fig. \ref{fig12f} highlighted the BAT region. These results demonstrate the feasibility of spatially separating $^{68}$Ga-PSMA and $^{18}$F-FDG distributions using the proposed triple-coincidence-based dual-isotope PET system, even under the low prompt gamma yield of $^{68}$Ga.

To quantitatively evaluate the separation performance, ROIs were drawn over the BAT and kidneys. As summarized in Table \ref{tab1}(c), the $^{68}$Ga-separated image retained 77.78\% of the kidney signal, with a residual signal ratio of 5.76\% in the BAT region. This low residual signal ratio suggests effective suppression of $^{18}$F-related contamination in the $^{68}$Ga-separated image, supporting the effectiveness of the CRT correction and isotope-separation workflow. In the $^{18}$F-separated image, 94.24\% of the BAT signal was preserved, whereas the residual signal ratio in the kidney region was 22.22\%. The higher residual signal ratio in the kidney region may be related to numerical fluctuations in the low-count $^{68}$Ga triple-coincidence image. These fluctuations can propagate through the image-domain subtraction process, leading to incomplete separation of the $^{68}$Ga component and residual signals in the $^{18}$F-separated image.

Overall, these findings validate the feasibility of $^{68}$Ga/$^{18}$F dual-isotope imaging utilizing weak triple coincidences from $^{68}$Ga and demonstrate the capability of the dual-isotope imaging system to restore isotope-specific uptake distributions \textit{in vivo}.

\FloatBarrier

\section{Discussion}
Efficient and accurate detection of triple-coincidence events is essential for simultaneous dual-isotope PET imaging based on prompt gamma emissions.
Most previously developed MI-PET systems rely on dedicated prompt gamma detectors to compensate for the intrinsically low detection efficiency and limited count statistics of triple-coincidence events.
However, such hardware-intensive designs inevitably increase system complexity and cost. Moreover, auxiliary prompt gamma detectors are typically positioned outside the main FOV, where prompt gamma detection efficiency may decrease with increasing distance from the center, leading to photon loss and reduced system efficiency. 
In contrast, the dual-isotope PET system presented in this work detects both annihilation photons and prompt gamma photons simultaneously using the same compact PET detector modules. By optimizing the geometric arrangement of standard PET panels, the proposed quad-panel system enhances detection efficiency across all photon types without requiring external prompt gamma detectors. Compared with the method proposed by Pratt et al.\cite{16}, which relies on hardware-based coincidence logic and V-shaped LOR, the present system leverages raw single-photon data and performs flexible multi-energy coincidence extraction entirely in software. This strategy preserves event-level energy and timing information and enables simultaneous extraction of double- and triple-coincidence events on the host workstation. Consequently, the system can support non-pure positron emitters with prompt gamma energies well above 511 keV, thereby expanding the range of radionuclide pairs that can be investigated for dual-isotope PET.

A notable feature of this work is the use of $^{68}$Ga as a non-pure positron emitter for dual-isotope imaging. $^{68}$Ga is clinically important but has a low prompt gamma emission probability of approximately 1.2\%, and its prompt gamma energy is much higher than 511 keV. Because of these challenges, $^{68}$Ga was often treated as a pure positron emitter in dual-tracer imaging studies. 
In this work, however, $^{68}$Ga was intentionally treated as a non-pure emitter to investigate the feasibility of simultaneous $^{18}$F/$^{68}$Ga imaging using weak triple-coincidence signals. 
System performance was evaluated using selected procedures adapted from NEMA NU 4-2008 and YY/T 1835-2022, followed by preclinical \textit{in vivo} validation. The results demonstrate that the system offers high spatial resolution, measurable triple-coincidence sensitivity, and effective separation of $^{68}$Ga/$^{18}$F tracer distributions.
These findings suggest that the proposed system is a practical preclinical platform for prompt-gamma-enabled dual-isotope PET, particularly for radionuclide combinations involving non-pure positron emitters with low prompt gamma yields and high-energy prompt gamma emissions.

The spatial-resolution results obtained with the $^{68}$Ga-filled Derenzo phantom should be interpreted in the context of radionuclide-dependent imaging performance rather than as an intrinsic detector resolution limit. The longer positron range of $^{68}$Ga compared with $^{18}$F contributes additional spatial blurring and can reduce the visibility of fine structures~\cite{Braune2022Resolution}. In a separate $^{18}$F Derenzo measurement using the same system, 1.2-mm rods were clearly resolved in the double-coincidence image (Fig.~S1). By comparison, the $^{68}$Ga double-coincidence image in Fig.~\ref{fig6} resolved 1.6-mm rods, whereas 1.2-mm rods were only partially distinguishable. This difference is consistent with positron-range effects, although count statistics and reconstruction settings may also contribute. For triple-coincidence imaging, spatial localization remains determined by the annihilation-photon pair. The reduced rod visibility in the $^{68}$Ga triple-coincidence image relative to the double-coincidence image may reflect limited triple-event statistics and the noise--resolution trade-off in reconstruction.

In addition to limited event statistics, random-triple contamination presents a further challenge for triple-coincidence imaging. In particular, CRTs are difficult to estimate accurately using conventional delayed-window methods.
As illustrated in Fig.\ref{fig8b}, even when the line-source region is restricted according to the sensitivity protocol, random-triple contamination remains non-negligible in the triple-coincidence data ($\sim20\%$) before correction. The low prompt gamma yield of $^{68}$Ga further reduces statistical reliability, making projection-domain pre-correction unstable and noise-sensitive.
To address this issue, we employed the consistent delayed expectation-maximization method for random-triple estimation and incorporated it into the reconstruction objective function. This approach allows true and random triple components to be estimated iteratively under a joint Poisson model. 

The effectiveness of this method was further validated using single-$^{18}$F mouse imaging data, for which true triple coincidences are theoretically absent. Under this condition, the detected triple-coincidence events are expected to be dominated by random triples. In a previous Monte Carlo study of the Argus scanner using a 3.40-MBq $^{18}$F source, CRTs accounted for 99.2\% of triple coincidences selected with all three photons within the 400--700-keV energy window~\cite{23}. 
As shown in Fig.~S2b, the conventional delayed-window correction left visible residual signals in the triple-coincidence image. In contrast, CDEM substantially suppressed these signals in the estimated true-triple image (Fig.~S2c), while the estimated random-triple image retained the predominant activity pattern (Fig.~S2d). These observations provide qualitative evidence of the effectiveness of CDEM in suppressing random-triple contamination under low-count conditions.

Low-count image reconstruction remains a major challenge for triple-coincidence imaging. In this study, we adopted a maximum-likelihood-based iterative reconstruction method. In the $^{124}$I/$^{18}$F dual-isotope mouse experiment, a uniform liver region, indicated by the red dashed ROI in Fig.\ref{fig11}, was used to calculate the signal-to-noise ratio (SNR)\cite{28}. 
For the $^{124}$I-separated image, the SNR decreased from 4.76 in the double-coincidence image of single $^{124}$I to 2.20 in the dual-tracer separation result. In contrast, the $^{18}$F-separated image did not show substantial SNR degradation. In the case of $^{68}$Ga/$^{18}$F dual-tracer imaging, the degradation was more pronounced. Because the prompt gamma yield of $^{68}$Ga is 1.2\%, and considering the finite prompt gamma detection efficiency of the system, the triple-to-double coincidence count ratio is approximately 1:1000. Using the liver region as background, we found that the SNR of the $^{68}$Ga-separated image dropped from 14.25 in double-coincidence image of single tracer to 4.58, while the SNR of the $^{18}$F-separated image decreased modestly from 10.24 to 8.28. These findings suggest that $^{68}$Ga triple-coincidence images are particularly susceptible to noise due to low event counts. To address this issue, future work will focus on improving low-count triple-coincidence reconstruction. One promising direction is to incorporate deep learning or learned regularization into iterative reconstruction frameworks to suppress noise while preserving anatomical structure and quantitative accuracy.

Another limitation of the current study is that isotope separation was performed in the image domain. In the present framework, the triple-coincidence image provides the distribution of the non-pure emitter, whereas the double-coincidence image represents a mixture of both isotopes. Isotope separation is then achieved by subtracting the scaled non-pure-emitter component from the double-coincidence image according to Eq. \ref{eqn_4}. Although this approach is straightforward and suitable for system performance evaluation, image-domain subtraction is sensitive to statistical fluctuations in low-count triple-coincidence images. These fluctuations may introduce negative values or residual artifacts in the separated images, thereby degrading image quality and affecting quantitative metrics. To overcome this limitation, we are investigating an advanced separation method based on joint Poisson modeling, in which the activity distributions of both isotopes and the random-triple component are estimated simultaneously during reconstruction. Such a model-based approach may reduce error propagation from low-count triple-coincidence reconstruction and improve the robustness of dual-isotope separation.

\section{Conclusions}
This study designed and evaluated a novel quad-panel dual-isotope PET system for simultaneous acquisition of double- and triple-coincidence events. By combining standard PET detector modules with GPU-accelerated software-based coincidence processing, the system enables flexible multi-energy event extraction from raw single-photon data without requiring dedicated prompt gamma detectors.
Phantom experiments demonstrated the system's capability for high-resolution imaging, prompt gamma detection, sensitivity measurement, quantitative activity assessment, and isotope separation fidelity. \textit{In vivo} experiments further confirmed the feasibility of separating $^{18}$F/$^{124}$I and $^{18}$F/$^{68}$Ga tracer distributions from single-session dual-isotope acquisitions. In particular, the successful separation of $^{68}$Ga/$^{18}$F demonstrates the potential of the proposed system for radionuclide combinations involving low prompt gamma yield and high-energy prompt gamma emission.
Overall, the proposed system provides a practical preclinical platform for prompt-gamma-enabled dual-isotope PET imaging. The use of standard detector modules and software-based coincidence processing may facilitate flexible multi-tracer imaging studies while reducing hardware complexity. Future improvements in low-count triple-coincidence reconstruction and model-based isotope separation are expected to further enhance image quality and quantitative robustness.

\backmatter

\section*{Abbreviations}

\begin{description}
    \item[PET] Positron emission tomography
    \item[CT] Computed tomography
    \item[MI-PET] Multi-isotope positron emission tomography
    \item[NETs] Neuroendocrine tumors
    \item[FDG] Fluorodeoxyglucose
    \item[FAPI] Fibroblast activation protein inhibitor
    \item[PSMA] Prostate-specific membrane antigen
    \item[DOTA] 1,4,7,10-Tetraazacyclododecane-1,4,7,10-tetraacetic acid
    \item[NaF] Sodium fluoride
    \item[LYSO] Lutetium-yttrium oxyorthosilicate
    \item[SiPM] Silicon photomultiplier
    \item[MVT] Multi-voltage threshold
    \item[BDM] Basic detector module
    \item[FOV] Field of view
    \item[LOR] Line of response
    \item[GPU] Graphics processing unit
    \item[EM] Expectation-maximization
    \item[MLEM] Maximum-likelihood expectation-maximization
    \item[TV] Total variation
    \item[CRT] Correlated random triple
    \item[CDEM] Consistent delayed expectation-maximization
    \item[SSRB] Single-slice rebinning
    \item[FWHM] Full width at half maximum
    \item[FWTM] Full width at tenth maximum
    \item[SF] Scatter fraction
    \item[NECR] Noise-equivalent count rate
    \item[ROI] Region of interest
    \item[SNR] Signal-to-noise ratio
    \item[BAT] Brown adipose tissue
    \item[NEMA] National Electrical Manufacturers Association
\end{description}

\FloatBarrier

\section*{Declarations}
\subsection*{Funding}
This work was supported in part by the National Natural Science Foundation of China (Nos. 62501415, 62027901, 62471372, and 62071362), the Natural Science Foundation of the Jiangsu Higher Education Institutions of China (No. 24KJD310001), the Natural Science Basic Research Program of Shaanxi (No. 2026JC-JCQN-093), the Jiangsu Government Scholarship for Overseas Studies, and the 2024 Qinglan Project of Jiangsu Province, China.

\subsection*{Competing interests}
All authors declare that they have no known conflicts of interest in terms of competing financial interests or personal relationships that could have an influence or are relevant to the work reported in this article.

\subsection*{Data availability}
The datasets used and/or analyzed during the current study are available from the corresponding author on reasonable request.

\subsection*{Ethics approval}
All animal experiments were conducted in accordance with the Ethical Guidelines on Animal Care and were approved by the Institutional Animal Care and Use Committee of Air Force Medical University (Fourth Military Medical University) with approval No. 251623.

\subsection*{Consent to participate}
Not applicable. 

\subsection*{Consent for publication}
Not applicable.

\subsection*{Authors' contributions}
YS and TJW contributed equally to this work. YS, FK and SPZ conceived and designed the study. YS, TJW, SY and BW developed the system workflow, including coincidence processing, correction, reconstruction and isotope separation. MN, YRW, RSY and FK designed and performed the animal experiments. YYL, WD and XYL contributed to system setup, data acquisition and experimental validation. YS, TJW, SY and BW analyzed and interpreted the phantom and animal imaging data. YS drafted the manuscript. SY, BW, FK and SPZ revised the manuscript critically for important intellectual content. FK and SPZ supervised the study and provided overall guidance. All authors read and approved the final manuscript.

\subsection*{Acknowledgements}
The authors thank Dr. Qingguo Xie for his valuable suggestions, and Raycan Technology Co., Ltd. (Suzhou) for providing the Trans-PET basic detector modules for dual-isotope PET.

\bibliography{ref}

\end{document}